\documentclass[a4paper,11pt]{article}
\usepackage{mathtools}
\usepackage{pgffor}
\usepackage{enumitem}
\usepackage{booktabs}
\usepackage{multirow}
\usepackage{color}
\usepackage{ulem}
\usepackage{listings}
\usepackage{}
\pdfoutput=1 

\usepackage{jheppub} 

\usepackage[T1]{fontenc} 
\usepackage{physics}
\usepackage{CJKutf8}
\usepackage[whole]{bxcjkjatype}

\usepackage[hang,small,bf]{caption}
\usepackage[subrefformat=parens]{subcaption}
\usepackage{tikz}
\usepackage{pgfplots} 
\usetikzlibrary{datavisualization}
\usetikzlibrary{datavisualization.formats.functions}
\usetikzlibrary{backgrounds}
\usetikzlibrary{patterns}
\usetikzlibrary{decorations.pathreplacing}
\usetikzlibrary{decorations.text}
\usetikzlibrary{decorations.markings}
\usetikzlibrary{decorations.pathmorphing}
\usetikzlibrary{trees}
\usetikzlibrary{quotes,angles}
\usetikzlibrary{shapes, calc, intersections, arrows, plotmarks, decorations, tikzmark, angles, quotes, babel, positioning}
\usetikzlibrary{arrows.meta}
\usepackage{multirow}
\usepackage{enumitem}

\renewcommand{\eqref}[1]{Eq.~\ref{#1}}
\newcommand{\figref}[1]{Figure \ref{#1}}
\newcommand{\tabref}[1]{Table \ref{#1}}
\newcommand{\secref}[1]{Section \ref{#1}}
\newcommand{\appref}[1]{Appendix \ref{#1}}

\newcommand{\mathrmbf}[1]{\mathrm{\mathbf{#1}}}

\newcommand{\ParT}{\mathrm{ParT}}
\newcommand{\AM}{\mathrm{AM}}
\newcommand{\FD}{\mathrm{FD}}
\newcommand{\PT}{\mathrm{P_T}}
\newcommand{\PQ}{\mathrm{P_Q}}
\newcommand{\HT}{\mathrm{H_T}}
\newcommand{\HQ}{\mathrm{H_Q}}

\newcommand{\MF}{\mathrm{MF}}
\newcommand{\kin}{\mathrm{kin}}

\newcommand{\subj}{\mathrm{subj}}
\newcommand{\high}{\mathrm{high}}
\newcommand{\low}{\mathrm{low}}
\newcommand{\PY}{\mathrm{PY}}
\newcommand{\HW}{\mathrm{HW}}

\title{\boldmath 
Reweighting and Analysing Event Generator Systematics by Neural Networks on High-Level Features
}
\author[a,b]{Amon Furuichi,\note{on leave to Sokendai. }}
\author[c, d]{Sung Hak Lim,}
\author[a, e]{Mihoko M. Nojiri}

\affiliation[a]{Graduate University for Advanced Studies (SOKENDAI) \\ 
Oho 1-1, Tsukuba, Ibaraki 305-0801, Japan}
\affiliation[b]{Department of Physics, Nagoya University \\
Furo-cho, Chikusa-ku, Nagoya, Aichi 464-8602 Japan}
\affiliation[c]{Particle Theory and Cosmology Group, Center for Theoretical Physics of the Universe, \\Institute for Basic Science (IBS), \\55 Expo-ro, Yuseong-gu, Daejeon 34126, Republic of Korea}
\affiliation[d]{Department of Physics and Astronomy, Rutgers, The State University of New Jersey, \\
136 Frelinghuysen Road, Piscataway, New Jersey 08854, USA}
\affiliation[e]{Theory Center, IPNS, KEK, \\
Oho 1-1, Tsukuba, Ibaraki 305-0801, Japan}

\emailAdd{amon@post.kek.jp}
\emailAdd{sunghak.lim@ibs.re.kr}
\emailAdd{nojiri@post.kek.jp}

\abstract{
The state-of-the-art deep learning (DL) models for jet classification use jet constituent information directly, improving performance tremendously. 
This draws attention to interpretability, namely, the decision-making process, correlations contributing to the classification, and high-level features (HLFs) representing the difference between signal and background. 
We address the interpretability issue using a modular architecture called the analysis model (AM), which combines several motivated HLFs as the input. 
We focus on the generator systematics of the top vs.~QCD classification by one of the best classifiers, Particle Transformer (ParT).  
Taking commonly used event generators \texttt{Pythia} (PY) and \texttt{Herwig} (HW) as examples, we demonstrate that the event weights estimated by the AM generator classifier align the HW classification score distribution to PY ones for QCD jets, with small training uncertainty. 
This suggests the AM is sufficient to describe simulated QCD jet features with relatively few observables, and generator systematics would also be reduced by reweighting the simulation by data. 
On the other hand, large event weights are required for QCD-like top jets, which leads to imperfect reweighting for both AM and ParT generator classifiers. 
Moreover, the AM HLFs are insufficient for describing PY and HW differences, causing lower reweighting accuracy compared with ParT.  
The missing features are the correlation among the collimated high-energy jet constituents, which are strongly correlated to the energy flow polynomials (EFPs) selected for top vs.~QCD classification,  
showing the complementarity between AM HLFs and the selected EFPs.
}

\preprint{CTPU-PTC-25-07}

\begin{document}
\maketitle
\flushbottom

\section{Introduction}

High-energy physics (HEP) experiments at the Large Hadron Collider (LHC) have fundamentally advanced our understanding of elementary particles, with the Higgs boson discovery marking a major milestone in modern particle physics.
One of the current objectives of HEP is precise measurements of Higgs properties, particularly its interactions, which are closely related to the stability of our vacuum, the beginning and fate of our Universe.
Any deviations from SM predictions would be hints for new physics, making these investigations crucial for current and future collider programs, including the High-Luminosity LHC (HL-LHC  \cite{ZurbanoFernandez:2020cco}), Future Circular Collider (FCC) \cite{Agapov:2022bhm}, and future Linear Colliders (ILC \cite{ILCInternationalDevelopmentTeam:2022izu}, CLIC \cite{Brunner:2022usy}).
In particular, measuring and understanding the behaviour of heavy SM particles such as top quark ($t$), $W$ and $Z$ bosons, and Higgs boson ($h$) is essential for probing signatures of physics beyond the Standard Model at the TeV scale.

A major progress in the study of heavy SM particles at LHC is in the tasks distinguishing boosted heavy SM particles from overwhelming QCD jet backgrounds. 
The boosted heavy particles exhibit unique kinematic and substructure features that can be used to distinguish them from QCD jets, but designing good classifiers remains a complicated task.
Recent advances in machine learning (ML), especially deep learning (DL), have provided powerful solutions by using jet substructures in-depth to achieve high classification accuracy \cite{PhysRevLett.121.241803, PhysRevD.101.075042, PhysRevD.107.016002, Qu:2022mxj, deOliveira:2015xxd, Farina:2018fyg, Kasieczka:2017nvn, Lin:2018cin, Lim:2020igi, Qu:2019gqs, Komiske:2018cqr, Dreyer:2020brq, Gong:2022lye}. 
DL methods are revolutionizing the analysis of HEP data as a whole. See recent developments in \cite{Kogler:2018hem, Larkoski:2017jix, Guest_2018, Radovic:2018dip}.

The DL models employ structured multilayer dense neural networks to model and learn complex correlations from large datasets. 
Among the known models,  Graph neural networks, such as ParticleNet \cite{Qu:2019gqs}, have demonstrated superior performance compared to image-based approaches \cite{Kasieczka:2019dbj}.
More recently, transformer architectures using self-attention mechanisms have proven their effectiveness in various tasks; Particle Transformer (ParT) \cite{Qu:2022mxj} has achieved high performance in jet classification.
These networks have further evolved to incorporate physics viewpoints.
For example, LundNet \cite{Dreyer:2020brq} captures information from jet clustering sequences, and PELICAN \cite{Bogatskiy:2022czk} embeds full 6-dimensional Lorentz symmetry and complete permutation equivariance. 
The cross-attention between subjets and jet constituents reduces the computational cost of the transformer networks significantly \cite{Hammad:2024cae}.

These new DL models in HEP often directly analyze low-level features (LLFs), such as the four-momenta of jet constituents, which provide primitive information for capturing complex correlations between the constituents.
While the direct use of LLFs has proven successful, it also presents significant practical and theoretical challenges.
LLFs are susceptible to low-energy (soft) particles, making them vulnerable to both experimental and theoretical systematic uncertainties.
On the theoretical side, predictions of parton showers and hadronization processes vary significantly among event generators like \texttt{Pythia} \cite{Bierlich:2022pfr}, \texttt{Herwig} \cite{Bellm:2019zci}, and Sherpa \cite{Bothmann_2019}, as well as Pythia extension to dipole-based shower models \cite{Fischer:2016vfv, Hoche:2015sya}. 
The discrepancies introduce systematic errors, particularly in modelling soft particle distributions.
Moreover, LLF-based models are prone to overfitting data noise and modelling artefacts of simulations, complicating their interpretation and reducing reliability.

The architectures using high-level features (HLFs) provide an alternative approach to achieving high performance in jet classification while mitigating the challenges of LLFs.
By focusing on collective jet properties rather than individual constituents, 
HLFs are less sensitive to noise and systematic uncertainties, providing greater stability and interoperability.
They are a baseline of robust and understandable classification while providing deeper insights into the underlying physics through interpretable quantities.

Leveraging the advantages of HLFs, a modular architecture called  Analysis Model (AM) \cite{Furuichi:2023vdx} was developed as a deep learning framework tailored for jet classification.
Its dedicated modules analyze subsets of HLFs, and their outputs are combined in a final classification module.
This modular design isolates individual HLF contributions, enhancing interpretability.
The AM balances classification accuracy and stability, providing a robust alternative to LLF-based approaches.

A set of HLF for AM has been proposed in \cite{Furuichi:2023vdx}, including global jet kinematics (e.g., transverse momentum, jet mass, and subjet momentum), two-point energy correlations which focus on energy correlation at a given angular distance \cite{Chakraborty:2020yfc}, and geometric measures of soft particles encoded through Minkowski functionals \cite{Chakraborty:2020yfc,Lim:2020igi}. 
The AM performs comparably with ParT for the top vs.~QCD classification with improved stability.

In addition, while it is not used for AM, energy flow polynomials (EFPs) are another useful high-level feature that characterizes jets by capturing permutation-invariant information through energy correlators \cite{Tkachov:1995kk} with polynomial angular weighting functions \cite{Komiske:2017aww,Kasieczka:2020yyl}. 
In \cite{Das:2022cjl}, it has been shown that the distance correlation (DisCo) method efficiently selects characteristics that contribute significantly to classification performance from a large pool of candidates for EFP.

AM can be used to adjust the simulated event distribution with experimental measurements through event reweighting \cite{Sugi:2012,Cranmer:2015bka, Brehmer:2018eca, Nachman:2020fff}. 
Current theoretical predictions based on event simulations often deviate from the experimental measurements due to insufficient higher-order QCD corrections and a poor understanding of non-perturbative QCD effects. 
The simulated distributions need to be adjusted by the actual data, while the required level of agreement between the real data and simulation is now higher than ever because DL models utilize subtle correlations among the event distributions. 
The DL methods themselves provide the solution to such problems. 
The output of the classifiers between the target data and simulated data can be used to build the event weights to align the simulated data distributions with the target data distributions. 
Reweighting simulated data using DL allows simultaneous adjustments of the multidimensional distributions. This is crucial for improving predictions and refining sensitivity estimates using DL. 
By utilizing the interpretability and robustness of HLFs, the reweighting using AM would allow the correction of simulations that are both physically tractable and less susceptible than the classifier using LLFs. 

To test this idea, we utilize AM to align predictions between \texttt{Pythia} (PY) and \texttt{Herwig} (HW), two commonly used event generators, by hypothesizing PY as target distributions. 
We evaluate the performance of HLF-based classifiers in reweighting the generator dependence away. 
We then test the quality of alignments by the score of signal vs.~background classification, to check if they contribute to reducing the generator systematics.

The paper is organized as follows.
\secref{section2} reviews the HLFs of AM and their implementation in the Analysis Model, followed by a comparison of event generator predictions for top vs.~QCD classification.
\secref{section3} introduces our methodology for constructing event weights to correct generator predictions. We apply this for the top vs.~QCD classification and show that AM achieves better precision than ParT in QCD jet reweighting, while ParT suffers from the variance in training. The stability of AM allows us to identify the relative importance of the HLFs in classification. On the other hand, both models face challenges in reweighting the top quark jets that contaminate into QCD jets (QCD-like top jets).  
\secref{section4} demonstrates that QCD-like top jets require significant reweighting factors, indicating substantial generator differences.
We also identify the limitations of the HLF sets of AM in characterizing top jet distributions and explore missing information through EFP distribution analysis; the remaining disagreement is described by the EFPs selected by the DisCo method. 
Finally, \secref{section5} discusses future directions. While parton shower prediction will likely improve in the near future with the NNLL accurate parton shower \cite{ Dasgupta:2018nvj,Karlberg:2021kwr,vanBeekveld:2023ivn}, simulation data still need to be adjusted by experimental data due to the incalculable non-perturbative QCD effects.
The HLFs used in this paper may bridge the gap between experimental data and theoretical predictions. 
Finally, \appref{subsec: boot} provides bootstrap-based error estimates for our results.

\section{High-Level Features describing Top Jet Classification at HCAL scale}
\label{section2}

Event simulations are the backbone of statistical analyses in collider physics, as they provide baseline predictions for reconstructed jet properties.
However, a key challenge arises from discrepancies between different event generators, each implementing different theoretical models and approximations.
Understanding these generator-specific differences and their impact on analysis results is crucial for reliable physics analyses.

This section compares top and QCD jet datasets generated by two widely used event generators, \texttt{Pythia} and \texttt{Herwig}.
\secref{sec: prepro_datasets} explains the details of our event simulation setup.
We note that the raw event datasets are those used in our previous paper \cite{Furuichi:2023vdx}.
In \secref{sec: classification_performance}, we review important high-level features (HLFs) describing top jet tagging. 
The selected features are used as inputs to a classifier model called the Analysis Model (AM) \cite{Furuichi:2023vdx}, which achieved performance comparable to the state-of-the-art Particle Transformer (ParT) \cite{Qu:2022mxj} when using features above the hadronic calorimeter (HCAL) angular scale of 0.1.
\secref{sec:dis_of_hlf_of_af} examines generator differences through selected HLFs.
Through this analysis, we identify features where generators require improvement and provide guidance for consistency between different generator predictions.

\subsection{Simulated Top and QCD Jet Datasets}
\label{sec: prepro_datasets} 

The top and QCD jet datasets from \texttt{Pythia} (PY) and \texttt{Herwig} (HW) are constructed as follows.
We first use \texttt{MadGraph5} 3.4.2 \cite{Alwall:2014hca} to generate events of the following parton-level processes for sampling jets: top-antitop pair production for top jets and dijet production for QCD jets.
The generated top quarks are decayed into three quarks using \texttt{madspin} \cite{Artoisenet:2012st}.
Simulated partons are then showered and hadronized using two different parton evolution simulators: \texttt{Pythia} 8.308 \cite{Bierlich:2022pfr} and \texttt{Herwig} 7.2 \cite{Bellm:2019zci}.
For the parton shower details, we employ "simple shower" for \texttt{Pythia} and the default shower for \texttt{Herwig}.

The generated particle-level events are then processed through the \texttt{Delphes} detector simulator \cite{deFavereau:2013fsa}.
We use a modified ATLAS detector card where jets are constructed from energy flow objects, which combine calorimeter energy deposits with tracking information to improve final-state particle reconstruction \cite{DeRoeck:2005sqi}.
All other detector simulation parameters follow the default ATLAS card.
Jets are then clustered using the anti-$k_T$ algorithm \cite{Cacciari:2008gp} with a radius parameter of $R = 1$, implemented in \texttt{fastjet} \cite{Cacciari:2011ma, Cacciari:2005hq}.

We select leading $p_{\mathrm{T}}$ jets satisfying transverse momentum $p_{\mathrm{T}}$  
$\in [500, \;600]\;$ $\mathrm{GeV}$, jet mass $m_\mathrmbf{J} \in [150,200]\;\mathrm{GeV}$, and pseudorapidity $|\eta| < 2$.
For top jet candidates, we additionally require all quarks from the top decay chain to lie within the distance $R=1$ from the jet axis in the pseudorapidity-azimuth ($\eta,\phi$) plane.
This parton matching requirement ensures that selected top jets contain all decay products from the top quark, excluding partially captured top jets from our analysis.

In this paper, we focus on the jet substructure at angular scales above HCAL granularity (0.1), where both the HLF-based Analysis Model and the LLF-based Particle Transformer show comparable classification performance \cite{Furuichi:2023vdx}.
At finer angular scales, the comparison becomes more complex due to additional systematic differences between PY and HW, along with detector-specific effects such as object type information from the tracker and electromagnetic calorimeter \cite{Park:2017rfb, Nakai:2020kuu}.
Since systematic differences exist even at the HCAL scale, we first focus on understanding and correcting these discrepancies at this coarser resolution.
The analysis of finer angular scales and the development of corresponding HLFs are left for future work.

For LLF analysis, we pixelate jets on the $(\eta,\phi)$ plane with pixel size $0.1\times0.1$ to integrate out sub-HCAL scale information.
When multiple jet constituents fall into the same pixel, their transverse momenta are summed.
These jet images are then regularized by following the procedure described in \cite{Lim:2020igi}.
The regularization aligns jets by placing the leading subjet at the origin, rotating the second subjet to lie along the positive $\phi$-axis ($\eta = 0, \phi > 0$), and ensuring the third subjet is positioned on the right side of the plane ($\eta > 0$) through mirroring.
This preprocessing ensures consistent jet orientation across samples, reducing the complexity of the network's learning task. 
A similar approach to masking sub-HCAL scale information is applied to each HLF, as described in the following subsection.

For clarity, we denote the datasets as follows:
\begin{itemize}
\item $\PT$ and $\HT$: top jets generated by PY and HW, respectively.
\item $\PQ$ and $\HQ$: QCD jets generated by PY and HW, respectively.
\end{itemize}
Each training and testing dataset contains approximately 0.5M samples for $\PT$ and 0.4M for $\PQ$, $\HT$, and $\HQ$.
Note that the measured cross-section for top pair production in this $p_{\mathrm{T}}$ range is about 300~fb \cite{ATLAS:2022xfj}.
These sample sizes correspond to an integrated luminosity of $\int \mathcal{L}\dd{t} \approx 1000\;\mathrm{fb}^{-1}$, providing statistical validity of our results at the High-Luminosity LHC (HL-LHC).

\subsection{Input Features of Particle Transformer and Analysis Model}
\label{sec: classification_performance}

This subsection compares two jet classifier models: Particle Transformer (ParT), which utilizes low-level features (LLFs) \cite{Qu:2022mxj}, and Analysis Model (AM), which leverages high-level features (HLFs) \cite{Furuichi:2023vdx}.
We present the mathematical definitions of relevant features \cite{Lim:2020igi, Chakraborty:2019imr}, describe the AM's modular architecture \cite{Furuichi:2023vdx}, and review their comparable performance in distinguishing top jets from QCD jets only using features above HCAL resolution scale.

\subsubsection{Low-Level Input Features of Particle Transformer}

Particle Transformer (ParT) is a state-of-the-art transformer-based jet classifier that processes LLFs, including energy, momentum, and particle identification of jet constituents \cite{Qu:2022mxj}.
In this study, we consider only kinematic features as we restrict our analysis to information above the HCAL angular scale.
Since particle identification requires tracker and electromagnetic calorimeter responses, we exclude this information from our analysis.
The detailed ParT setup used in this paper is equivalent to that in \cite{Furuichi:2023vdx}. Especially, given that our sample size is about 400k per class, we reduce the ParT parameters down to 340k, by reducing the number of attention blocks and the width of each neural network from the default ParT setting.

Within our HCAL-scale analysis, the constituent-wise kinematic input features for ParT are as follows.
\begin{itemize}
\item $\Delta\eta$: difference in pseudorapidity $\eta$ between the particle and the jet axis.
\item $\Delta\phi$: difference in azimuthal angle $\phi$ between the particle and the jet axis.
\item $\log p_{\mathrm{T}}$: logarithm of the particle's transverse momentum $p_{\mathrm{T}}$.
\item $\log E$: logarithm of the particle's energy.
\item $\log(p_{\mathrm{T}}/p_{\mathrm{T},\mathrmbf{J}})$: logarithm of the particle's transverse momentum $p_{\mathrm{T}}$ relative to the jet transverse momentum $p_{\mathrm{T},\mathrmbf{J}}$.
\item $\log(E/E_{\mathrmbf{J}})$: logarithm of the particle's energy relative to the jet energy $E_{\mathrmbf{J}}$.
\item $\Delta R = \sqrt{(\Delta \eta)^2 + (\Delta \phi)^2}$: angular separation between the particle and the jet axis on pseudorapidity-azimuth $(\eta, \phi)$ plane.
\end{itemize}

In addition, ParT uses augmented attention blocks with pairwise particle information, motivated by QCD interactions \cite{Dreyer:2020brq}.
The additional features for particle pairs $(a,b)$ are as follows.
\begin{itemize}
\item $\Delta = \sqrt{(\Delta y)^2 + (\Delta \phi)^2}$: angular separation between particles $a$ and $b$ on rapidity-azimuth $(y, \phi)$ plane.
\item $k_T = \min(p_{\mathrm{T},a}, p_{\mathrm{T},b}) \Delta$: $k_T$ distance \cite{Ellis:1993tq}.
\item $z = \min(p_{\mathrm{T},a}, p_{\mathrm{T},b})/(p_{\mathrm{T},a}+p_{\mathrm{T},b})$: momentum fraction of the softer constituent.
\item $m$: invariant mass of the two-particle system.
\end{itemize}
Explicitly considering these variables allows the attention block to directly process important information in jet physics.

\subsubsection{High-Level Input Features of Analysis Model}
\label{sec:hlf_am}
\begin{figure}[ht]
    \centering
    \begin{subfigure}[b]{0.4\textwidth}
        \centering
\includegraphics[width=\textwidth]{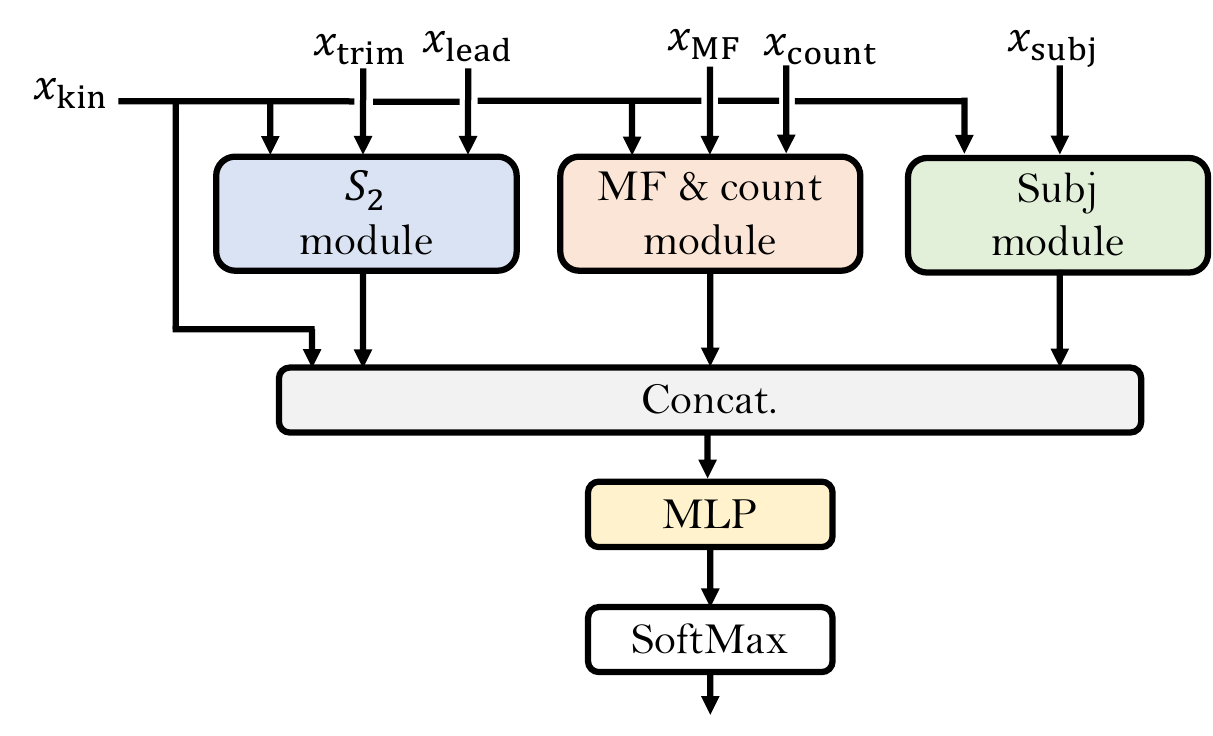}
        \caption{ }
        \label{fig:all}
    \end{subfigure}
    \begin{subfigure}[b]{0.4\textwidth}
        \centering
\includegraphics[width=\textwidth]{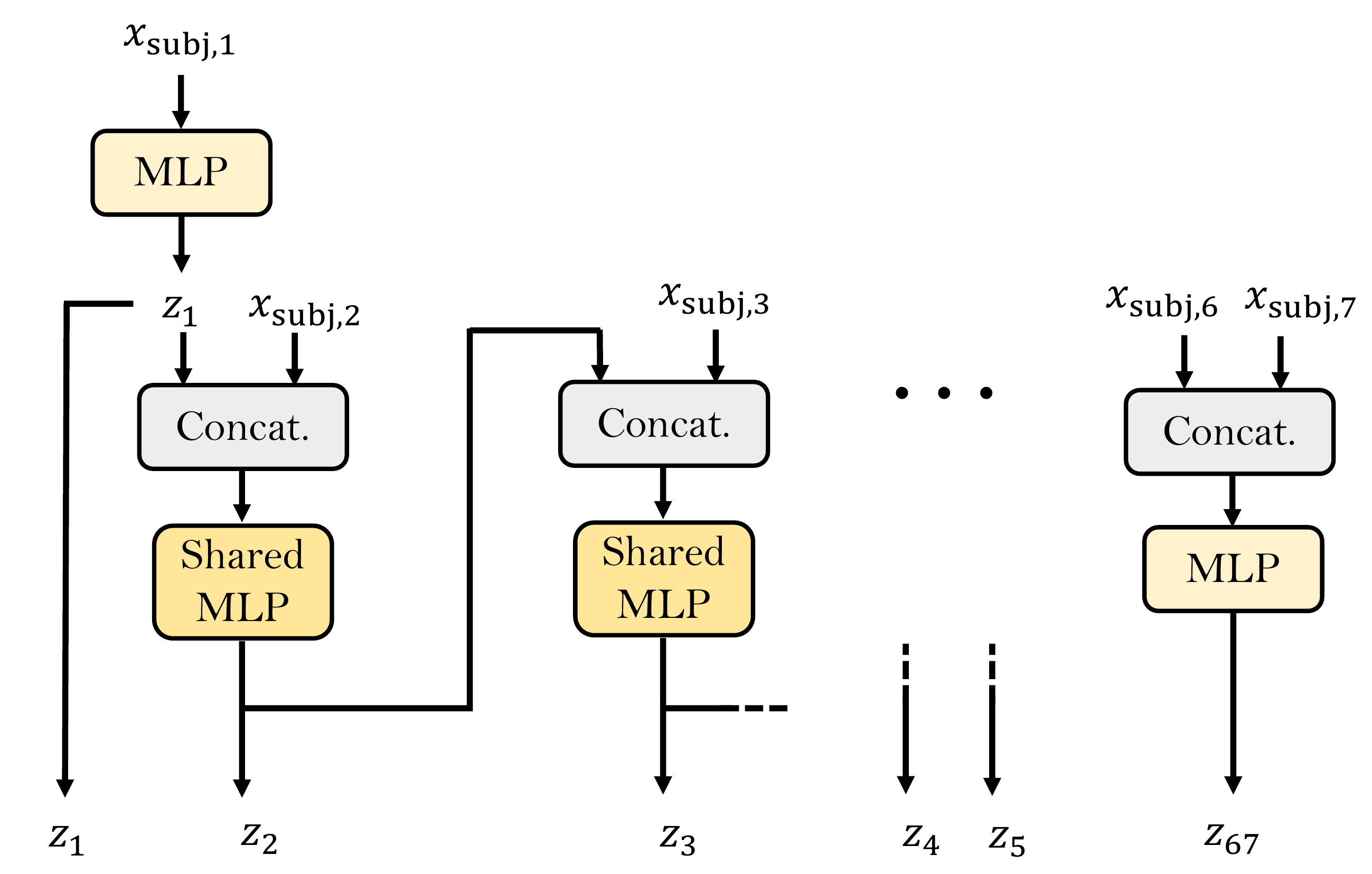}
        \caption{}
        \label{fig: subjmodul}
    \end{subfigure}
    \caption{
   Structure of Analysis Model (AM).  (a) The global structure of the network.   $S_2$,  $\MF$ + count, and subjet module outputs are concatenated to go to the final MLP. $S_2$, $MF$+ count module are simple MLP. 
     (b) Structure of Subjet module. The n-th subjet is  processed by shared MLP with (n-1)-th subjet module output. The figure is taken from \cite{Furuichi:2023vdx}.}
    \label{fig:AMstr}
\end{figure}

Analysis Model (AM) is a modular architecture in which each module analyzes predefined HLFs important in jet tagging. 
Here, we use the same setup described in \cite{Furuichi:2023vdx}.
The structure of the network is illustrated in Fig. \ref{fig:AMstr}.
AM includes three specialized modules: the $S_2$ module for energy correlations \cite{Lim:2020igi}, the $\MF$ module for jet morphology and constituent multiplicities \cite{Chakraborty:2019imr}, and the subjet module for analyzing colour substructure of subjets.
The body of the $S_2$ and $\MF$ modules is MLP, while the subjet module is using a recursive network as shown in \figref{fig:AMstr}(b). Each MLP module typically has hidden layer with 100 or 200 with depth of two layers, with total numbers of trainable parameters of 300k, comparable to the size of the training dataset.
We briefly describe these HLFs used in each of the modules below.

\paragraph{Global Kinematical Features ($x_\mathrm{kin}$)}
To provide information about the mass and energy scales of clusters inside a jet to each module, we include the following kinematical features:
\begin{align}
x_\mathrm{kin}
=
(p_{\mathrm{T},\mathrmbf{J}}, m_\mathrmbf{J}, p_{\mathrm{T},\mathrmbf{J}_\mathrm{trim}}, m_{\mathrmbf{J}_\mathrm{trim}}, p_{\mathrm{T},\mathrmbf{J}_\mathrm{lead}}, m_{\mathrmbf{J}_\mathrm{lead}}).
\end{align}
The trailing subscripts of $p_{\mathrm{T}}$ and $m$ denote the following sub-clusters considered for given variable:
\begin{itemize}
\item $\mathrmbf{J}$: entire jet.
\item $\mathrmbf{J}_\mathrm{trim}$: trimmed jet, obtained using the $k_T$ algorithm \cite{Catani:1993hr, Ellis:1993tq} with a radius of 0.2, retaining subjets with energy fractions above 0.05.
\item $\mathrmbf{J}_\mathrm{lead}$: leading $p_{\mathrm{T}}$ subjet, defined as the highest $p_{\mathrm{T}}$ anti-$k_T$ subjet within a radius of 0.2 \cite{Cacciari:2008gp}.
\end{itemize}

\paragraph{$S_2$ Module: Two-Point Energy Correlations}
The $S_2$ module processes two-point energy correlation spectra \cite{Lim:2018toa, Chakraborty:2019imr} between clusters.
For clusters $a$ and $b$, the correlation $S_{2,ab}(R)$ at angular scale $R$ is defined as:
\begin{align}
S_{2,ab}(R)
\coloneqq
\sum_{i \in a, j \in b} p_{\mathrm{T},i} p_{\mathrm{T},j} \delta(R - R_{ij}),
\end{align}
where $R_{ij}$ is the angular separation between jet constituents $i$ and $j$.
This function $S_2(R)$ is infrared and collinear (IRC) safe and forms a basis for general two-point energy correlations, allowing this module to quantify both the jet's prong structure and correlations between hard and soft radiation.

\paragraph{MF Module: Generalized Jet Constituent Multiplicity}
This module analyzes Minkowski Functionals (MFs), which quantify the geometric and topological characteristics of jet constituent distributions on the $(\eta,\phi)$ plane \cite{Lim:2020igi, Chakraborty:2020yfc}.
The MFs for two-dimensional data analysis consist of three functionals:
\begin{itemize}
\item area ($A$)
\item perimeter length ($L$)
\item Euler characteristic ($\chi$)
\end{itemize}
These quantities are computed after dilating each constituent on the $(\eta,\phi)$ plane with a circle of radius $R$ \cite{Lim:2020igi}.
The MFs form a basis set for geometric measures, including jet constituent multiplicities, which are represented by Euler characteristics at a dilation scale of $R=0$.
Thus, the MFs generalize constituent multiplicities, an important variable for analyzing the colour charge of the originating parton at the leading order.
Moreover, MFs as functions of $R$ encode detailed geometric and topological information.
When dilation from scale $R_1$ to $R_2$ induces only simple geometric expansion of isolated patches, 
the MFs at larger scales are determined entirely by those at smaller scales.
Deviations from this standard dilation behaviour indicate non-trivial structures at the corresponding scale.

To incorporate $p_{\mathrm{T}}$ information, we compute MFs for different subsets of jet constituents, each set including only constituents above a given $p_{\mathrm{T}}$ threshold up to 8GeV.
Additionally, this module analyzes jet constituent $p_{\mathrm{T}}$ histograms to directly probe constituent multiplicity at different energy scales.
These additions enhance the jet tagging performance of this MF module \cite{Furuichi:2023vdx}.

\paragraph{Subjet Module: Subjet Color Analysis}
Quarks from top decay form the backbone of the substructure of the top jet, while gluons are the primary source of QCD jet substructures.
Therefore, subjet colour information provides additional discrimination power.
To serve this purpose, this subjet module analyzes up to seven leading subjets using recursive neural networks (RNNs) \cite{Furuichi:2023vdx}.
For each subjet, we consider the following features:
\begin{itemize}
\item transverse momentum ($p_{\mathrm{T}}$)
\item spatial coordinates ($\eta, \phi$)
\item mass ($m$)
\item constituent multiplicity ($N_c$)
\end{itemize}
In our HCAL-scale analysis, this set of constituent multiplicity and kinematic features provides sufficient enhancement to the tagger performance.

\subsubsection{Classification Performance Comparison}
We now briefly review the classification performance of ParT and AM as presented in \cite{Furuichi:2023vdx}.
We evaluate performance using the area under the ROC curve (AUC) and rejection rates $R_{X\%}$ at truepositive rates of $X\%$.
The classification metrics AUC, $R_{50\%}$, and $R_{30\%}$ for both models on PY and HW datasets are summarized in \tabref{tab: cla}.
From this comparison, we observe:
\begin{enumerate}
\item ParT and AM achieve comparable classification performance, demonstrating their effectiveness in distinguishing top jets from QCD jets with information limited to HCAL resolution.
\item Differences in results between PY and HW highlight non-negligible differences between samples generated by those simulators.
\end{enumerate}
While ParT effectively captures complex, high-dimensional correlations through LLFs, our previous work \cite{Furuichi:2023vdx} also demonstrated that AM achieves comparable performance with less training uncertainty in prediction because AM is a more constrained model than ParT. 
This feature makes AM particularly suitable for reweighting as it provides more reliability. 
Furthermore, AM provides enhanced interpretability through its use of physics-motivated HLFs.
This interpretability of AM inputs enables the quality assessment of reweighted outputs by examining HLFs that capture key jet substructure characteristics.

While AM and ParT performance is similar, the computational requirements are very different. 
During training with a batch size of 1,000, AM requires less than 1 GB of GPU memory, compared with 14 GB of ParT.
The ratio of the number of floating point operations is approximately 9,  indicating that AM requires fewer GPU operations than ParT \cite{Furuichi:2023vdx}.

\begin{table}
\begin{center}
\begin{tabular}{l|cc|cc}
\toprule
& 
\multicolumn{2}{c|}{$\PT$ vs.~$\PQ$} & 
\multicolumn{2}{c}{$\HT$ vs.~$\HQ$} 
\\
\midrule
Model & ParT & AM & ParT & AM
\\
\midrule
AUC & 
0.944 & 0.943 &
0.929 & 0.928
\\
$R_{50\%}$ & 
90.5 & 85.7 &
62.6 & 61.3
\\
$R_{30\%}$ & 
372 & 343 &
242 & 244
\\
\bottomrule
\end{tabular}
\end{center}
\caption{
Classification metrics of the top jet tagging by ParT and AM. 
$\PT$ vs.~$\PQ$ is the results using \texttt{Pythia}-simulated jets.
$\HT$ vs.~$\HQ$ is the results using \texttt{Herwig}-simulated jets. The numbers are from \cite{Furuichi:2023vdx}.
}
\label{tab: cla}
\end{table}

\subsection{Distributions of AM's High-Level Features}
\label{sec:dis_of_hlf_of_af}

This subsection examines distributions of high-level features (HLFs) used in AM for top and QCD jet classification.
We briefly compare a few selected HLF distributions from PY and HW generators to understand how the simulator difference appears on each distribution.

\begin{table}
\begin{center}
\begin{tabular}{lcccc}
\toprule
& $\PT$ 
& $\HT$ 
& $\PQ$ 
& $\HQ$ 
\\
\midrule
multiplicity & 
35.7 &
35.9 &
43.4 &
41.9 
\\
$S_2$[$10^3$ GeV$^2$]& 
28.8 &
28.6 &
14.3 &
14.5 
\\
MFs (number change)  & 
0.800 &
0.808 &
0.960 &
0.925 
\\
\bottomrule
\end{tabular}
\end{center}
\caption{
Mean values of the distributions in \figref{fig: diff_all}.
The mean values are well-tuned, while we have seen a large deviation -away from those values in \figref{fig: diff_all}. 
}
\label{tab: ave}
\end{table}

\begin{figure}[t]
    \centering
    \begin{subfigure}[b]{0.3\textwidth}
        \centering
        \includegraphics[width=\textwidth]{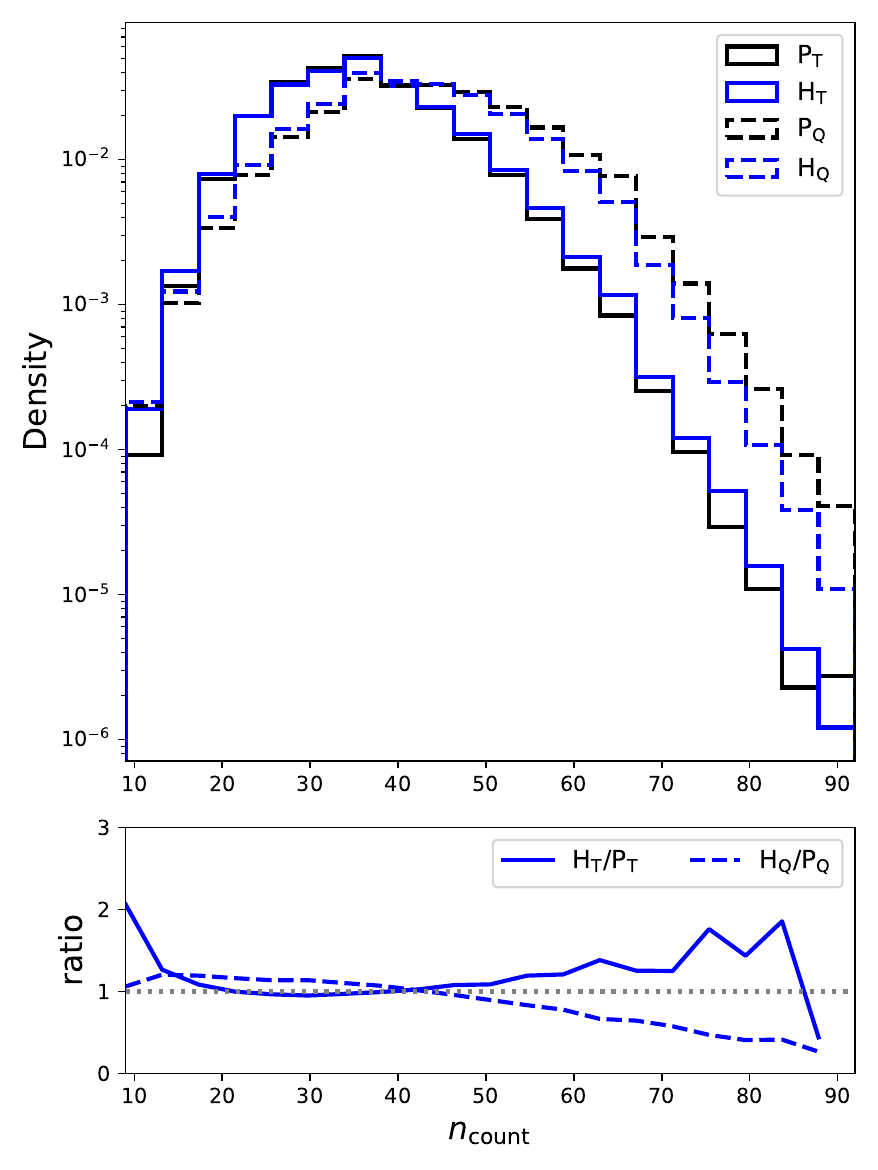}
        \caption{constituent multiplicity}
        \label{fig: diff_num}
    \end{subfigure}
    \begin{subfigure}[b]{0.3\textwidth}
        \centering
        \includegraphics[width=\textwidth]{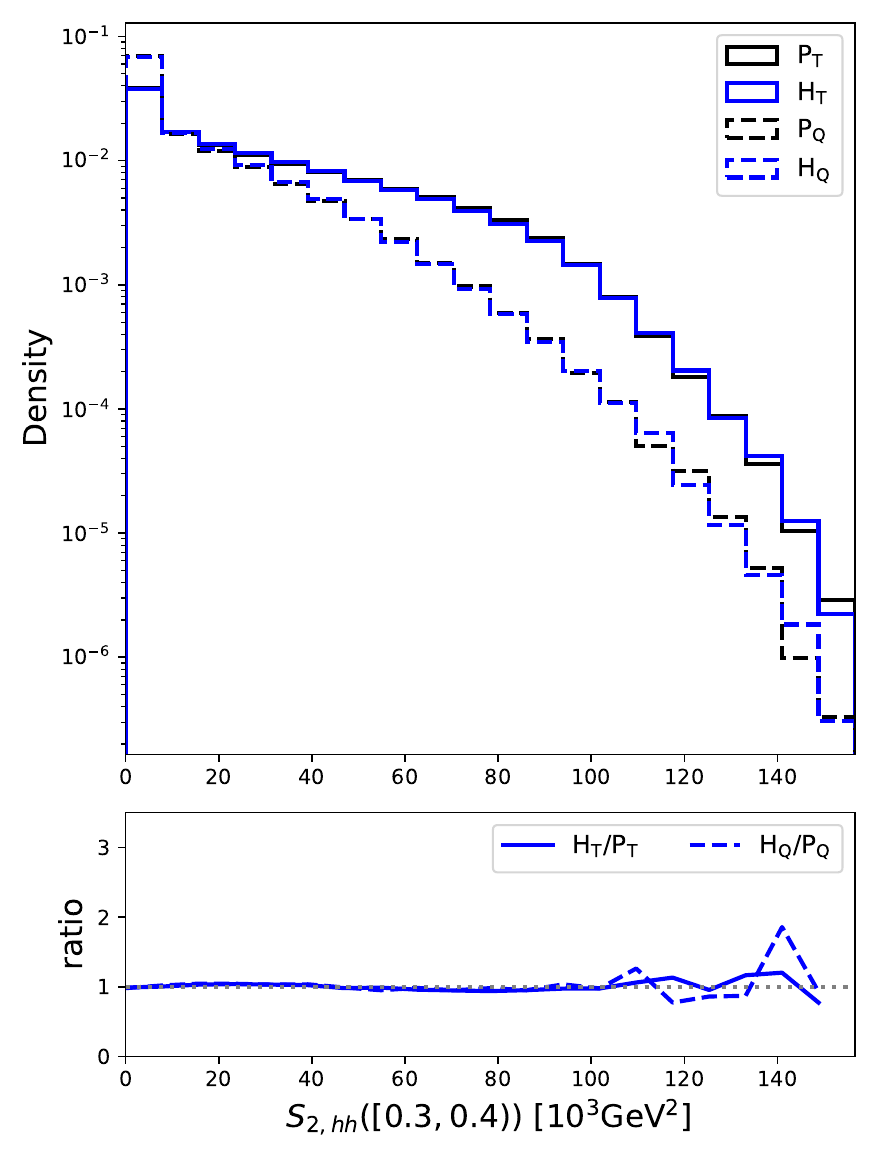}
        \caption{$S_2$ in $R=[0.3-0.4)$}
        \label{fig: diff_s2}
    \end{subfigure}
    \begin{subfigure}[b]{0.3\textwidth}
        \centering
        \includegraphics[width=\textwidth]{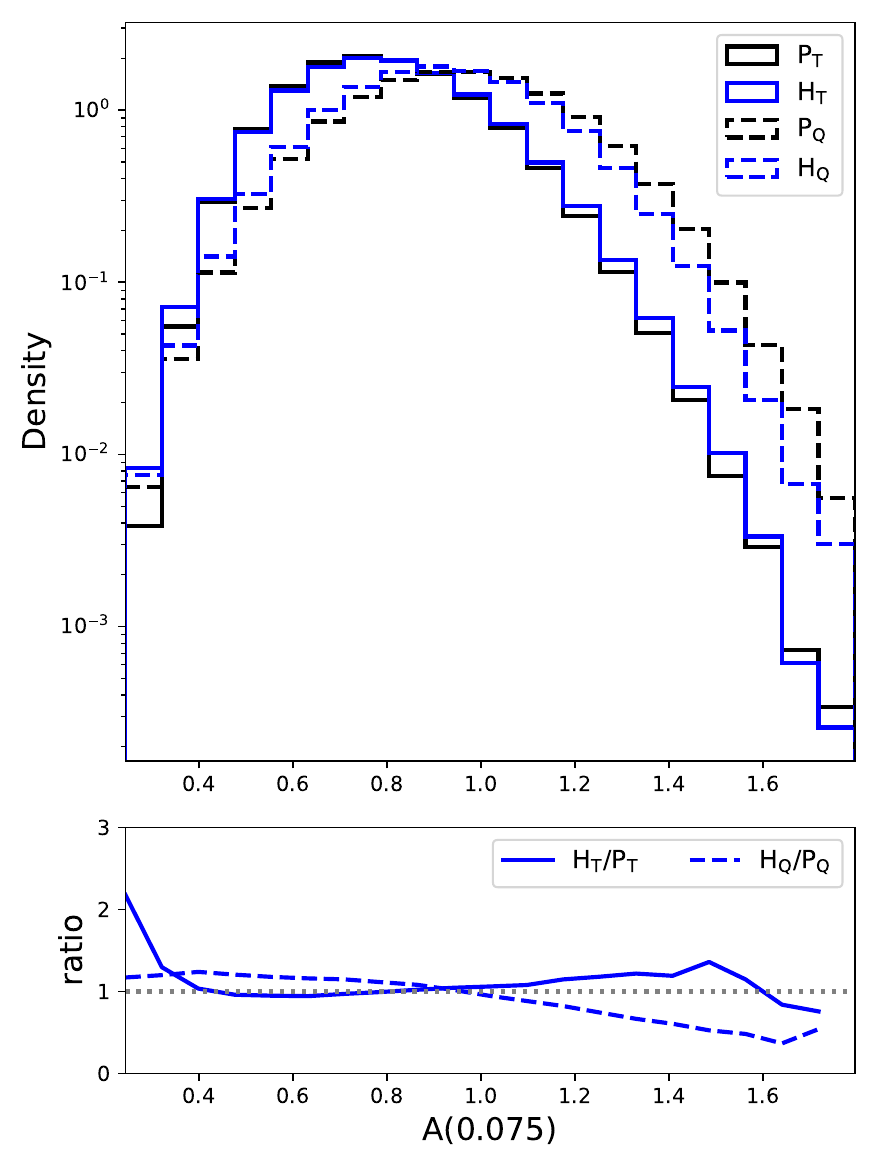}
        \caption{MF: area with $R=0.075$}
        \label{fig: diff_mf}
    \end{subfigure}
    \caption{
    Top: The normalized histograms of HLFs used as AM inputs for the top (solid line) and QCD (dashed line) samples. Bottom: The ratio to PY distribution corresponding to the top plot. The resolution is set to $\Delta R=0.1$. a) constituent multiplicity of the image binned by $\Delta R=0.1$, b) $S_2$ for $R=[0.3-0.4)$ c)  MF: area $A$ of the dilated image with $R= 0.075$. }
    \label{fig: diff_all}
\end{figure}

\paragraph{Constituent Multiplicity}
Constituent multiplicity ($n_\mathrm{const}$) is a variable that often shows significant differences between Monte Carlo simulations, as it is not an IRC-safe observable and cannot be described perturbatively under QCD radiation. 
The distribution is directly affected by details of parton shower and hadronization modeling.
\figref{fig: diff_num} shows constituent multiplicity histograms for top and QCD jets.
Solid and dashed lines represent top and QCD jets, respectively, with black lines showing \texttt{Pythia} ($\PT$, $\PQ$) and blue lines showing \texttt{Herwig} ($\HT$, $\HQ$) results.
\tabref{tab: ave} shows that the mean constituent multiplicities are well-matched across simulators.

The differences between simulated top jets are relatively small overall since top jets consist primarily of quark subjets, which are well-tuned across various Monte Carlo simulations.
However, differences still appear in the high-multiplicity tail of the distribution.
The $\HT$ and $\PT$ show opposite behavior as $n_\mathrm{const}$ increase, 
$\HT/\PT=1.25$ at $n_\mathrm{const} = 70$, 
while $\HQ/\PQ=0.62$. 
These events with high multiplicity correspond to more radiative processes with additional QCD radiation, where simulator dependence increases due to less-tuned gluon jet modelling.

For QCD jets, the differences between simulators become apparent for a larger fraction of events than for top jets.
This is expected as QCD jets in our dijet sample are predominantly gluon-initiated.
The greater simulator dependence of gluon jets is more readily visible here, as these jets are directly initiated by gluons rather than arising from heavy particle decay and subsequent parton showers. 
Such differences at the tail of the distribution impact classification metrics.

\paragraph{Two-Point Energy Correlation Spectrum ($S_2$)}
The two-point energy correlation $S_2$ measures energy correlations between constituent pairs at specific angular separations.
In \figref{fig: diff_s2}, we show the $S_2$ distribution of trimmed jets for angular scale window $R\in[0.3,0.4)$, which corresponds to the characteristic subjet opening angle for the top jets in our analysis.
\tabref{tab: ave} summarizes the mean $S_2$ values for top and QCD jets.
The simulator dependency is relatively low because $S_2$ is an IRC-safe functional with subjet kinematics governed mainly by original parton kinematics. 
The distributions show better agreement between simulators than the constituent multiplicity case.

\paragraph{Area in Minkowski Functionals (MFs)}
Minkowski Functionals encode the geometric and topological properties of jet constituent distributions \cite{Lim:2020igi, Chakraborty:2020yfc}.
In \figref{fig: diff_mf}, we show the histogram of one MF: the area at a dilation scale of $R=0.075$, where separated pixels begin to merge as the diameter $2R$ exceeds the HCAL angular resolution of 0.1.
\tabref{tab: ave} lists the mean values of this variable.

Although the mean values are well aligned, significant deviations occur away from the mean.
The differences between simulators shown in \figref{fig: diff_mf} follow a similar pattern to the constituent multiplicity histogram in \figref{fig: diff_num}, as the area function is proportional to constituent multiplicity at small $R$.
This dilation scale of 0.075 is smaller than the characteristic angular separation between quarks from top decay ($\sim 0.3$).
Consequently, this MF distribution remains sensitive to parton shower and hadronization model details.
As MFs are IRC-unsafe observables, these geometric measures are particularly susceptible to differences in soft radiation and underlying event modelling.

\section{Calibrating Simulated Events using Deep-Learned Reweighting Factor}
\label{section3}
\subsection{Reweighting Function Estimation using Classifiers}

Simulated events for real data analysis are often calibrated using reweighting techniques to align approximate simulations to observed data.
A conventional approach is reweighting based on discrete histograms of a few selected important features.
However, this method can only refine a limited number of features simultaneously as histograms experience difficulty in modelling data on a high-dimensional space.

To consider many features at once, reweighting based on neural networks and density ratio estimation (see \cite{Sugi:2012} for a review) is gaining attention, with examples including \cite{Cranmer:2015bka, Brehmer:2018eca, Andreassen:2019nnm, Nachman:2020fff}.
These neural network-based techniques utilize classifiers that estimate posterior probabilities or density ratios to compute the reweighting factor $w_{\mathrm{T}/\mathrm{G}}(x)$ from a simulation $\mathrm{G}$ to the target $\mathrm{T}$:
\begin{equation}
w_{\mathrm{T}/\mathrm{G}}(x) = \frac{p(x|\mathrm{T})}{p(x|\mathrm{G})},
\end{equation}
where $p(x|\mathrm{T})$ and $p(x|\mathrm{G})$ are the probability density functions of features $x$ for the target $\mathrm{T}$ and simulation $\mathrm{G}$, respectively.

This weight function corrects the dataset $\mathrm{G}$ to match $\mathrm{T}$ through importance sampling.
The expected value of an observable $f(x)$ for a target distribution $\mathrm{T}$ can be evaluated by the weighted sum of dataset $\mathrm{G}$ with weights $w_{\mathrm{T}/\mathrm{G}}(x)$:
\begin{equation}
\int \dd{x} p(x|\mathrm{T}) \cdot f(x) =
\int \dd{x} p(x|\mathrm{G}) \cdot \frac{p(x|\mathrm{T})}{p(x|\mathrm{G})} \cdot f(x)
\approx
\frac{1}{N} \sum_{i=1}^N  w_{\mathrm{T}/\mathrm{G}}(x^{(i)}) \cdot f(x^{(i)}),
\end{equation}
where the superscript $(i)$ denotes features $x$ from the $i$-th sample among $N$ samples.

For high-dimensional data analysis such as jet classification, neural network-based classifiers offer a scalable solution by estimating density ratios via posterior probability $p(\mathrm{T}|x)$.
When we consider equal class probabilities $p(\mathrm{T})=p(\mathrm{G})=0.5$, the weighting function $w_{\mathrm{T}/\mathrm{G}}(x)$ relates to the posterior probability $p(\mathrm{T}|x)$ as:
\begin{equation}
{w}_{\mathrm{T}/\mathrm{G}}(x) = \frac{p(\mathrm{T}|x)}{1 - p(\mathrm{T}|x)}.
\end{equation}
This formula allows us to convert the neural network output $s(x)$ into the estimated weight function $\tilde{w}_{\mathrm{T}/\mathrm{G}}(x)$, as training with binary cross-entropy loss function makes $s(x)$ estimate $p(\mathrm{T}|x)$.
The estimated weight $\tilde{w}_{\mathrm{T}/\mathrm{G}}(x)$ is then defined as follows:
\begin{equation}
\tilde{w}_{\mathrm{T}/\mathrm{G}}=\frac{s(x)}
{1-s(x)}. 
\end{equation}
While reweighting is theoretically robust in the asymptotic limit ($N\rightarrow \infty$), the performance is affected by two practical factors:
\begin{itemize}
\item \textbf{Statistical Uncertainties}: Limited sample sizes introduce uncertainties in weight estimation.
\item \textbf{Inductive Biases}: Classifiers may fail to approximate posterior probabilities due to model assumptions, architecture limitations, and hyperparameter choices.
\end{itemize}
Therefore, to reweight generated samples accurately and precisely, we require careful control of both statistical and systematic uncertainties.
In the following, we will demonstrate that AM fulfills these demands for simulated events prepared for top jet tagging. 

\subsection{Benchmark 1: Statistical Fluctuation of Normalization during Reweighting}

The event reweighting based on a finite dataset suffers from statistical fluctuations.
These effects become prominent in expressive networks, as their flexibility demands larger training samples, potentially leading to larger uncertainties in reweighted distributions.

This statistical effect in reweighting can be demonstrated in the sample mean of the weights \cite{nachman2020neural, PhysRevD.98.052004},
\begin{equation}
    \langle \tilde{w}_{\mathrm{T}/\mathrm{G}} \rangle = \frac{1}{N} \sum_{i=1}^{N} \tilde{w}_{\mathrm{T}/\mathrm{G}}(x^{(i)}). 
\end{equation}
In $N\rightarrow \infty$ limit, the expectation value converges to $\langle \tilde{w}_{\mathrm{T}/\mathrm{G}} \rangle \rightarrow 1$  as  $\int\dd{x}p(x|\mathrm{G}) \tilde{w}_{\mathrm{T}/\mathrm{G}}(x) = \int\dd{x}p(x|\mathrm{T}) = 1$, i.e., event normalization is preserved.
Therefore, any deviation from 1  can be interpreted as error propagation from classifier modelling of reweighting.

\begin{table}[ht]
\centering
\begin{tabular}{r|cc|cc}
\toprule
Dataset     & \multicolumn{2}{c|}{QCD jets} & \multicolumn{2}{c}{top jets} \\
Reweighting & $\HW \rightarrow \PY$ & $\PY \rightarrow \HW$ & $\HW \rightarrow \PY$ & $\PY \rightarrow \HW$ \\
\midrule
ParT & $\phantom{+}3.61 \pm 2.23\%$ & $-1.62 \pm 2.07\%$ & $\phantom{+}3.79 \pm 2.72\%$ &$-0.56 \pm 2.42\%$ \\
AM & $-0.38 \pm 0.30\%$ & $\phantom{+}0.29 \pm 0.26\%$ & $\phantom{+}0.36 \pm 0.37\%$ & $-0.49 \pm 0.43\%$ \\
\bottomrule
\end{tabular}
\caption{
Deviations of the weight function expectation $\langle \tilde{w}_{\mathrm{T}/\mathrm{G}} \rangle$ from 1 (in percent).
Centre values and standard deviations are estimated from multiple network training with different random seeds.}
\label{tab:norm}
\end{table}

The sample mean of the weights shows larger deviations from 1 for ParT.  \tabref{tab:norm} summarizes the deviations of $\langle \tilde{w}_{\mathrm{T}/\mathrm{G}} \rangle$ for HW $\rightarrow$ PY and PY $\rightarrow$ HW reweightings.
AM demonstrates better normalization control, with deviations $\langle \tilde{w}_{\mathrm{T}/\mathrm{G}} \rangle-1$ consistently below 0.5\%.
In contrast, ParT exhibits deviations of several percent, larger than those of AM by almost an order of magnitude.
These results indicate that even larger training datasets would be needed to achieve sub-percent normalization precision 
with ParT. Note that the classifiers are trained using samples with size about $5\times 10^5$, which is comparable to the number of events at HL-LHC in the same $p_{\mathrm{T}}$ range, namely the available target dataset at most.

Nevertheless, this normalization bias can be corrected by rescaling the weight function $\tilde{w}_{\mathrm{T}/\mathrm{G}}$.\footnote{For other classifier calibration methods, see \cite{guo2017calibrationmodernneuralnetworks}. The calibration methods may be applicable to further improve the weight function estimation with a proper tranform from the calibrated classifier output to the probability density ratio.}
We define the rescaled weight function $\tilde{w}_{\mathrm{T}/\mathrm{G}}$ as follows:
\begin{align}
{w}_{\mathrm{T}/\mathrm{G}} (x) 
= \frac{\tilde{w}_{\mathrm{T}/\mathrm{G}}(x)}{\langle \tilde{w}_{\mathrm{T}/\mathrm{G}} \rangle}.
\end{align}
The rescaled weight ensures $\langle {w}_{\mathrm{T}/\mathrm{G}} \rangle = 1$ by construction, making it more suitable for a detailed comparison between reweightings using ParT and AM.

\begin{figure}[ht]
\centering
\includegraphics[width=0.45\textwidth]{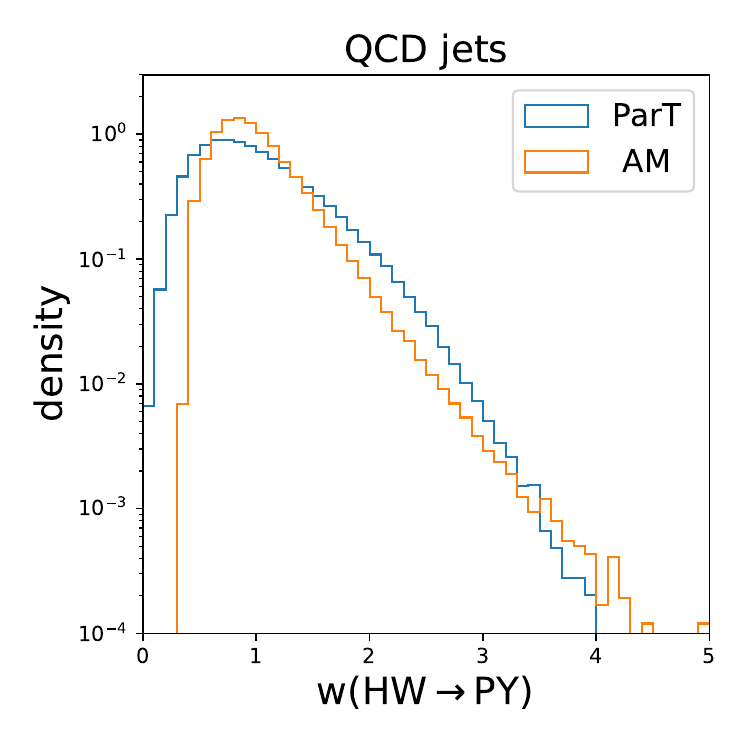}
\includegraphics[width=0.45\textwidth]{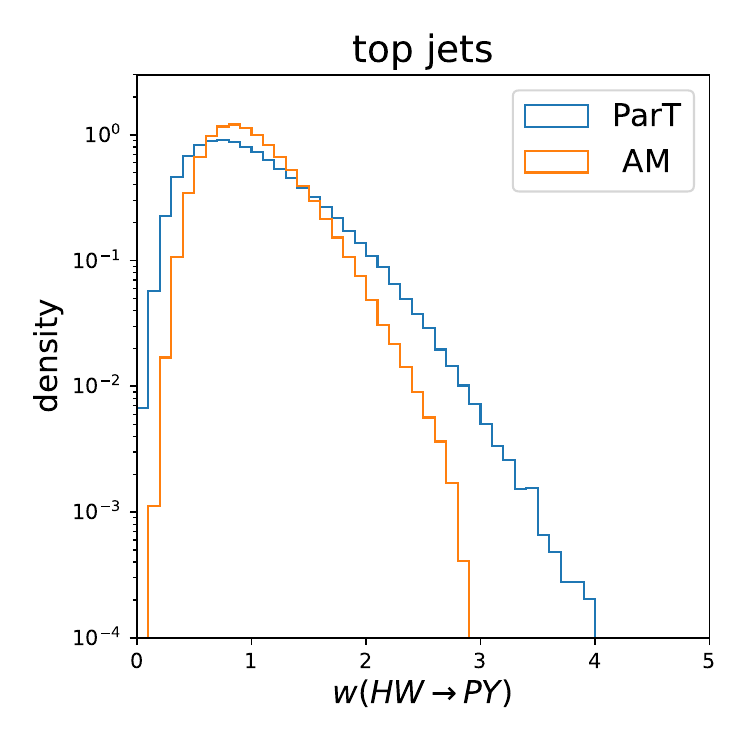}
\caption{
Distributions of rescaled weights ${w}_{\mathrm{T}/\mathrm{G}}$ for HW $\rightarrow$ PY reweighting. 
The left panel shows QCD jets, and the right panel shows the top jets. 
Blue histograms are for ParT, and orange histograms are for AM.
}
\label{fig:weight}
\end{figure}

\figref{fig:weight} shows the distributions of rescaled weights in HW $\rightarrow$ PY reweightings.
ParT produces broader weight distributions, indicating more aggressive corrections with weights deviating further from 1.
This explains the larger statistical fluctuations in ParT-based reweighting, as extreme weights affect the reweighted distributions: large weights amplify statistical uncertainties by emphasizing high-weight samples, while small weights effectively remove samples from the analysis.

In contrast, AM yields narrower weight distributions, leading to reduced statistical uncertainties through milder corrections.
However, this stability comes with a trade-off: the difference in rescaled weight distributions suggests that AM's HLF set may not capture some critical information needed for reweighting HW-generated jets to PY-generated jets in rare event regions.
The choice between architectures for reweighting thus depends on task-specific requirements for precision versus accuracy.

\subsection{Benchmark 2: Assessing Reweighting Quality with Top Jet Classifiers}

\begin{figure}[ht!]
    \begin{center}
        \begin{tikzpicture}[scale=1.0]
            \node [
                draw,
                rectangle,
                minimum height=1.0cm,
                minimum width=4cm,
                outer sep=0.25cm
            ] 
            (top_g1) at (0,3) {HW-simulated top jets $(\HT)$};
            \node[
                draw,
                rectangle,
                minimum height=1.0cm,
                minimum width=4cm,
                outer sep=0.25cm
            ] 
            (qcd_g1) at (8,3) {HW-simulated QCD jets $(\HQ)$};
            \node [
                draw,
                rectangle,
                minimum height=1.0cm,
                minimum width=4cm,
                outer sep=0.25cm
            ] 
            (top_g2) at (0,0) {PY-simulated top jets $(\PT)$};
            \node [
                draw,
                rectangle,
                minimum height=1.0cm,
                minimum width=4cm,
                outer sep=0.25cm
            ] 
            (qcd_g2) at (8,0) {PY-simulated QCD jets $(\PQ)$};
            \draw [
                -{Latex[width=0.25cm]}
            ] (qcd_g1) 
                edge node[right, align=left] {1. reweight $\HQ$ using \\\phantom{1. }ParT/AM for \\ \phantom{1. } ($\HQ$ vs. $\PQ$)} 
            (qcd_g2);
            \draw [
                -{Latex[width=0.25cm]}
            ] (top_g1) 
                edge node[right, align=left] {1. reweight $\HT$ using \\ \phantom{1. }ParT/AM for \\ \phantom{1. } ($\HT$ vs. $\PT$)} 
            (top_g2);
            \draw [
                {Latex[width=0.25cm]}-{Latex[width=0.25cm]}
            ] (qcd_g2) 
                edge node[below=0.65, align=left] {2. test reweighing quality \\  \ using "test classifier" \\ \ $\ParT$    for ($\PT$ vs. $\PQ$)} 
            (top_g2);
        \end{tikzpicture}
    \end{center}
    \caption{ 
    A flow chart of the reweighting quality test using a top jet classifier. We first reweight HW-simulated datasets to PY-generated datasets using ParT or AM, and we test the reweighting quality using a "test classifier" ParT trained on tagging PY-generated top jets.  
    }
    \label{fig:qualtet}
\end{figure}
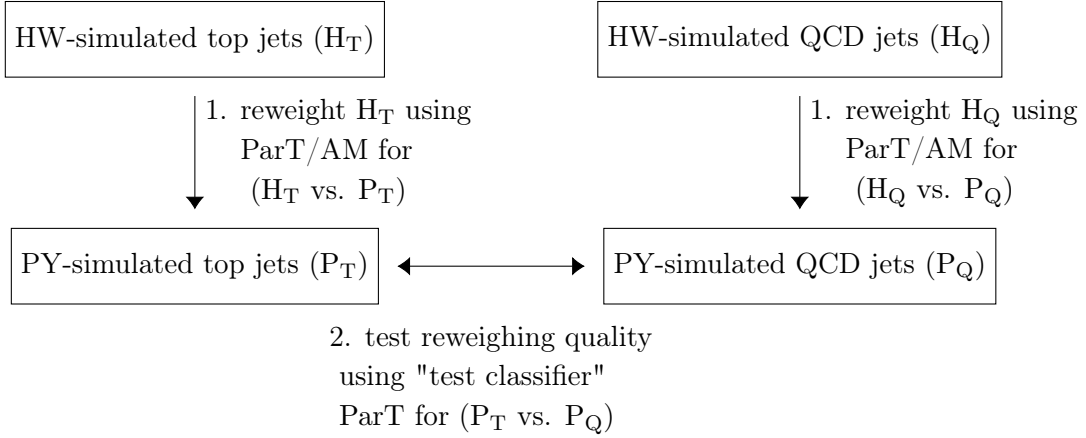

To assess reweighting quality in the context of top jet tagging, we examine whether ParT and AM-based reweighting correctly calibrates the features needed for the classification.
We reweight HW-generated jets to match PY-generated jets and evaluate whether the reweighted dataset produces classifier output distributions compatible with the original PY-generated dataset. 
We use ParT trained for the $\PT$ vs.~$\PQ$ as the "test classifier" since ParT considers a variety of features through its LLF-based approach.
If the reweighting is sufficient, the ParT output distribution for reweighted HW-generated jets should align with that of the PY-generated sample.
We illustrate the flow chart of our strategy in \figref{fig:qualtet}.

\begin{figure}[ht]
    \centering
    \includegraphics[width=0.5\textwidth]{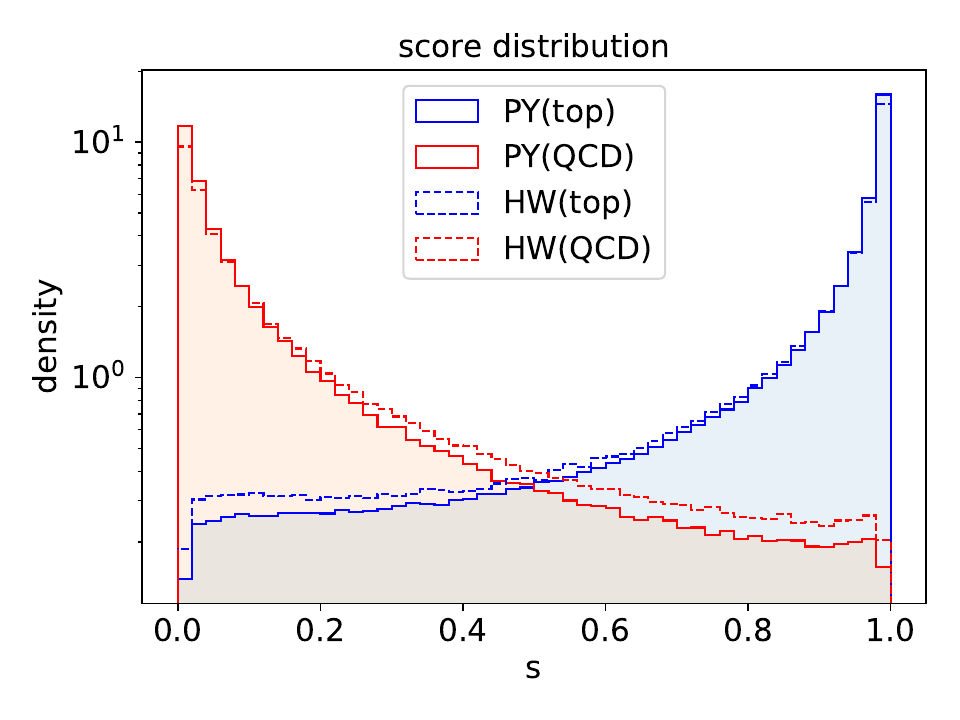}
    \caption{The normalized score distributions of the classifier trained for $\PT$ vs.~$\PQ$ datasets. The score distributions are calculated for QCD jets (red) and top jets (blue) generated by PY (solid) and HW (dotted).  
    }
    \label{fig:scoredis}
\end{figure}

\figref{fig:scoredis} shows the distributions of the outputs of  ParT-based top jet classifier output distributions for PY and HW samples. 
The shapes of distributions are similar, but we observe a sizable difference in the misclassified region, specifically in the high (low) classifier output region for QCD (top) jets, respectively. 
To quantify this difference, we examine the jet rejection performance $R_{X\%}$, which measures the classifier's ability to suppress background jets at a given signal efficiency $X\%$. 
For the test based on PY samples, the performance is high, with $R_{50\%}=90.5$ and $R_{30\%}=372$. 
In contrast, the test based on HW samples shows lower performance, with $R_{50\%}=71.7$ and $R_{30\%}=281$, indicating notable generator-dependent variations in classifier output distributions.

These discrepancies are mainly due to the difference in subleading QCD interaction modelling in simulations.
For example, PY and HW adopt different parton shower schemes. 
HW adopts an angular-ordered shower model characterized by wide-angle and soft particle emissions. 
PY employs a $p_{\mathrm{T}}$-ordered shower model with angular ordering veto, which produces different particle emission patterns. 
These differences make the modelling of wide-angle emissions in PY and HW different, influencing the classifier output distributions.

The quality of reweighting can be further analyzed through the normalized histogram ratio of classifier output distributions.
The normalized histogram ratio in the interval of $s=[s_1,s_2]$ is 
\begin{equation}
r(s) \coloneqq \frac{N(s\vert \HW; w)}{N(s \vert \PY)},
\label{eq:rwratio}
\end{equation}
where $N(s \vert \HW; w)$ represents the fraction of the weight sum for HW events in the interval $s$ reweighted by the $w$, and the weight sum for all the HW events.
$N(s \vert \PY)$ is the fraction of bare PY events in the interval $s$. 
The weight parameter $w$ is suppressed in the left hand side.
This ratio will converge to 1 if the reweighting is perfect.

In the narrow bin width limit, the histogram ratio for reweighted HW jets over PY jets corresponds to the following density ratio at the bin center $s$,
\begin{equation}
r(s) = \frac{p(s|\mathrm{HW} \rightarrow \mathrm{PY})}{p(s|\mathrm{PY})},
\end{equation}
where $p(s|G)$ is the density of the test classifier output $s$ for a dataset $G$. $\mathrm{HW} \rightarrow \mathrm{PY}$ denotes the reweighted HW dataset.

\begin{figure}[ht!]
 \begin{minipage}[t]{0.48\linewidth}
\centering
\includegraphics[width=\textwidth]{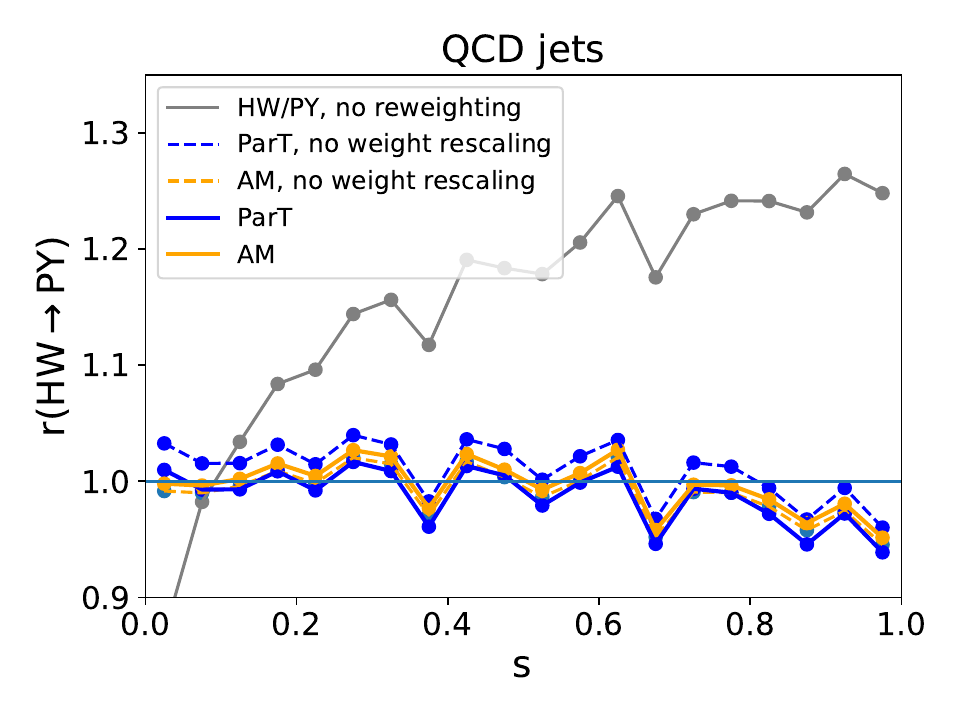}
\subcaption{The ratio $r(\HW\rightarrow \PY)$ for QCD jets}
\end{minipage}
 \begin{minipage}[t]{0.48\linewidth}
\includegraphics[width=\textwidth]{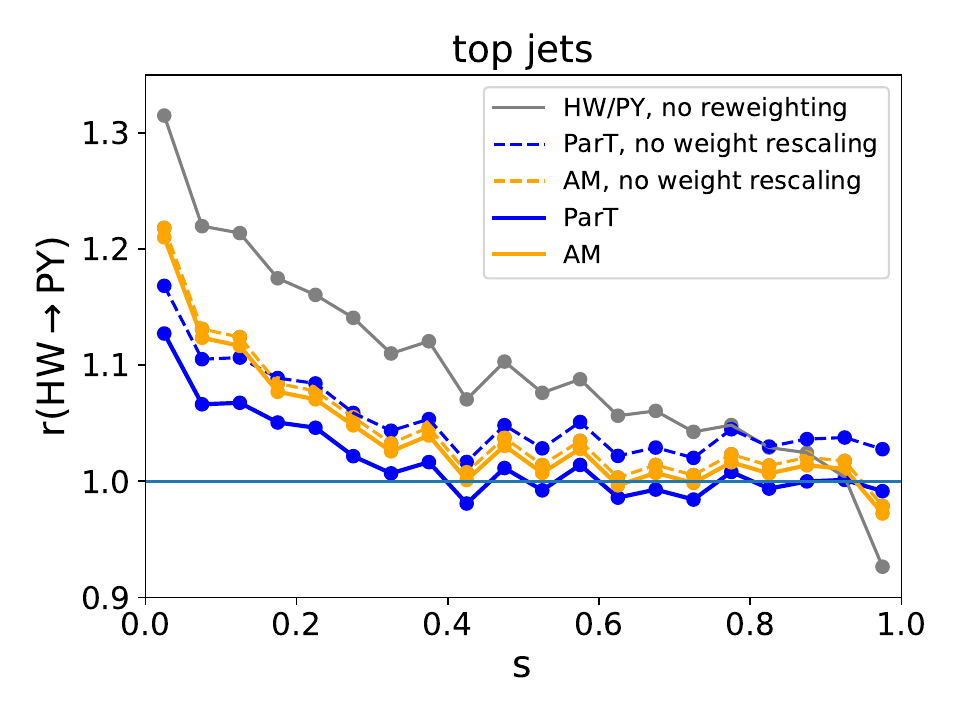}
\subcaption{The ratio $r(\HW\rightarrow \PY)$ for top jets}
\end{minipage}
\caption{
The normalized histogram ratios of the reweighted HW distribution to the PY distribution $r(\HW\rightarrow\PY)=r(s)$ for the ParT output $s$ trained for $\PT$ vs.~$\PQ$. (a) is for QCD jets $\PQ$ and $\HQ$, and (b) is for top jets $\PT$ and $\HT$.
The blue and orange lines are the ratios using ParT and AM, respectively. 
The dashed lines are the results of reweighting without weight scaling, namely using $\tilde{w}$, and the solid lines are the ones using the rescaled weight $w$. The ratio will be close to one if the corresponding reweighting model accurately calibrates the HW dataset into the PY dataset. 
The grey lines are the ratio between HW and PY jets without using reweighting.
}
\label{fig:score_reweight}
\end{figure}

\figref{fig:score_reweight}(a) and \ref{fig:score_reweight}(b) illustrate the histogram ratios of QCD and top jets, respectively, for ParT (blue lines) and AM (orange lines).
The reweighting results with (without) weight rescaling are shown by solid lines (dashed lines).
A perfect reweighting would result in a ratio of one for each bin.
All reweighted distributions show correlated bin-wise fluctuations arising from the statistical fluctuation of the training and testing datasets.
The training errors are estimated using the bootstrapping method detailed in \appref{subsec: boot}.
The weight rescaling is important for ParT;  the ParT reweighting without weight rescaling (blue dashed lines) gives the worst reweighting accuracy for both QCD and top jets.

For QCD jets, $r(s)$ is consistent with one across the entire classifier output region.
This indicates that both AM and ParT can effectively correct generator systematics.
Given the smaller statistical uncertainties of AM compared to ParT \cite{Furuichi:2023vdx}, reweighting using AM is advantageous, providing similar results with a simpler, more interpretable network based 
on HLFs.

The situation is different for top jets.
First of all, compared with the QCD jet, the difference between $\PT$ and $\HT$ shown in the grey line is somewhat small for top-like jets $s\sim 1 $, while it is very large for QCD-like top jets at $s\sim 1$. 
Top-like jets have a multi-prong structure associated with quarks coming from top decay. 
The quark fragmentations in PY and HW are tuned to the same LEP data, suppressing the systematic differences around $s\sim 1$.  
The deviations of the ratio from one before and after reweighting are prominent in QCD-like regions with small $s$ as shown in \figref{fig:re_sratio}(b). All reweighting results by ParT and AM 
show significant deviations from 1 in the QCD-like region with $s \sim 0$. Moreover, ParT-based reweighting is better than AM.  
We will show in \secref{sec:results} that the deviation comes from missing specific features that the HLFs of AM do not cover.

\begin{table}[ht!]
\begin{center}
\begin{tabular}{c|cc|cc}
\toprule
\multirow{ 2}{*}{QCD jets}  &   \multicolumn{2}{c|}{QCD-like region} & \multicolumn{2}{c}{top-like region}   \\
   &   \multicolumn{2}{c|}{$s<0.05$} & \multicolumn{2}{c}{$s \in (0.5,0.7)$}   \\
\midrule
weight &  rescaled & no scaling & rescaled & no scaling \\
\midrule
ParT  & $\phantom{+}0.010$ & $\phantom{+}0.033$  & $-0.016$ & $\phantom{+}0.006$ \\
AM    & $-0.007$ & $-0.007$ & $\phantom{+}0.003$ & $\phantom{+}0.004$ \\
\bottomrule
\bottomrule
\multirow{ 2}{*}{top jets}  &   \multicolumn{2}{c|}{top-like region} & \multicolumn{2}{c}{QCD-like region}  \\
  &   \multicolumn{2}{c|}{$s >0.95$} & \multicolumn{2}{c}{$s \in (0.3,0.5)$}   \\
\midrule
weight &  rescaled & no scaling & rescaled & no scaling \\
\midrule
ParT  & $-0.009$ & $\phantom{+}0.028$   &   $\phantom{+}0.003$ & $\phantom{+}0.041$ \\
AM    & $-0.019$ & $-0.016$  &    $\phantom{+}0.016$ & $\phantom{+}0.015$ \\
\bottomrule
\end{tabular}
\end{center}
\caption{ 
Deviations of normalized weight sum ratio in \eqref{eq:rwratio} from one, $r(s)-1$, for QCD and top jets in QCD-like region and top-like region.
The definition of the range $s$ is on the first row of each table.
We present two results using rescaled weights $w$ and bare weights $\tilde{w}$.
}
\label{tab:pamratio}
\end{table}

\tabref{tab:pamratio} shows the numerical values of the deviation of the ratio from one, $r(s)-1$  for QCD jets $\PQ$ and $\HQ$ in QCD-like region ($s < 0.05$) and moderately top-like region ($s\in(0.5,0.7)$) and for top jets $\PT$ and $\HT$ in top-like region ($s>0.95$) and moderately QCD-like region ($s\in (0.3, 0.5)$).
The numerical values confirm poor alignment of ParT without weight rescaling, good alignment using AM for the QCD jets, and relatively poor alignment using AM for the top jets.

In conclusion, for the number of training samples of about 0.5M, ParT faces statistical stability issues, but the problem is reduced after correcting the overall normalization.  
ParT and AM demonstrate good reweighting accuracy for QCD jets, but AM shows superior stability and reweighting accuracy, particularly for QCD jets.  
On the other hand, both ParT and AM face difficulty finding the proper weight for reweighting the QCD-like top events. 
The deviation of the ratio from 1 is around 0.1 for ParT and 0.2 for AM. 
It is plausible that the models cannot focus on classification in this region due to the reduced statistics. 
We also find AM is underperforming in this region; the origin of the gap in the performance will be discussed in \secref{sec:results}.

\subsection{Importance of High-Level Features of Analysis Model in Reweighting}

In this subsection,  we investigate the impact of each HLF of AM in reweighting by utilizing only parts of AM modules in the analysis. 
We denote such reduced AM as AM-PIPs (Partial InPuts) \cite{Furuichi:2023vdx}.
Each AM module is designed to capture specific jet substructures as described in \secref{sec:hlf_am}; we could assess the importance of each HLF in reweighting through the performances of the AM-PIPs.  

For this analysis, we consider the following three versions of AM-PIPs by systematically using different groups of substructure inputs. 
Note that $x_{\kin}$ is included for all the AM-PIP setups to set mass and energy scales.
\begin{itemize}
    \item MF(low): the jet constituent $p_{\mathrm{T}}$ histogram bins below $p_{\mathrm{T}} \approx 4.4\;\mathrm{GeV}$, and Minkowski functionals with low $p_{\mathrm{T}}$ thresholds of 
    $p_{\mathrm{T}} > 0.5, 1, 2 \;\mathrm{GeV}$),
    \item MF(high): the jet constituent $p_{\mathrm{T}}$ histogram bins above $p_{\mathrm{T}} \approx 4.4\;\mathrm{GeV}$, and Minkowski functionals with high-$p_{\mathrm{T}}$ constituent thresholds  ($p_{\mathrm{T}} > 4, 8 \;\mathrm{GeV}$).
\end{itemize}
Here, MF(low) focuses on geometric features and multiplicities involving low $p_{\mathrm{T}}$ jet constituents, while MF(high) mainly considers high $p_{\mathrm{T}}$ jet constituents.
The union of MF(high) and MF(low) is 
the inputs of the MF modules. 
\begin{itemize}
    \item $S_2 + \mathrm{subj}$: the two-point energy correlation module and subjet module mainly using subjet kinematics information though the input also includes a constituent multiplicity of the subjets. 
\end{itemize}
This setup focuses on the IRC-safe features of the AM modules, making it suitable for testing how far the AM-based reweighting can reach by using IRC-safe features.

\begin{figure}[htbp]
 \begin{minipage}[t]{0.48\linewidth}
 \centering
    \includegraphics[width=\textwidth]{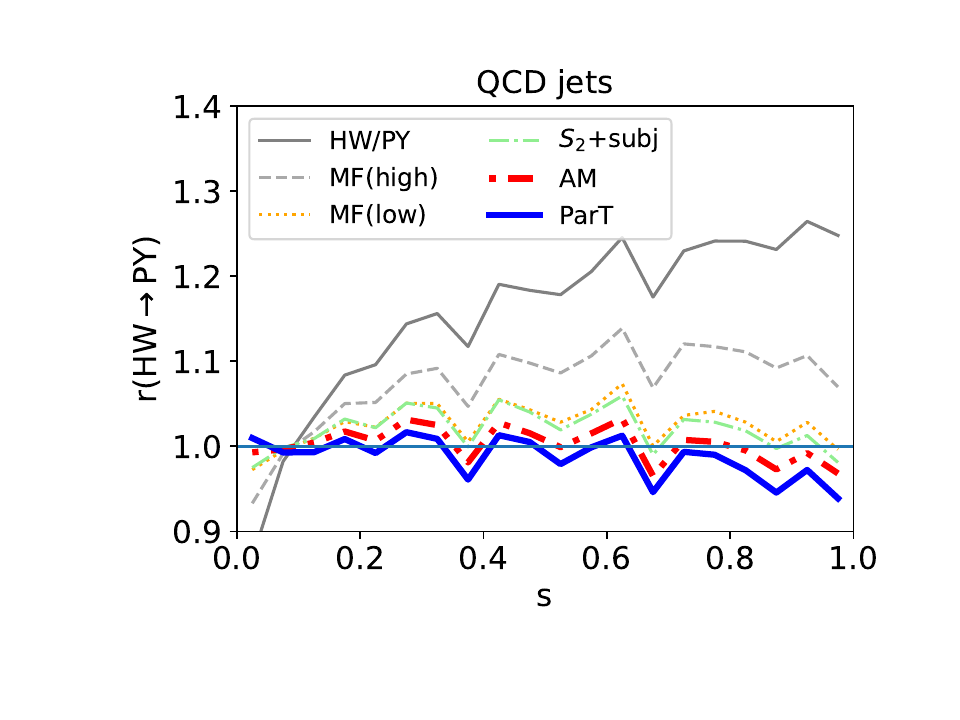}
\vskip -0.5cm
    \subcaption{ $r( \HW\rightarrow \PY)$ for QCD jets $\PQ$ and $\HQ$}
    \end{minipage}
\begin{minipage}[t]{0.48\linewidth}
\centering
    \includegraphics[width=\textwidth]{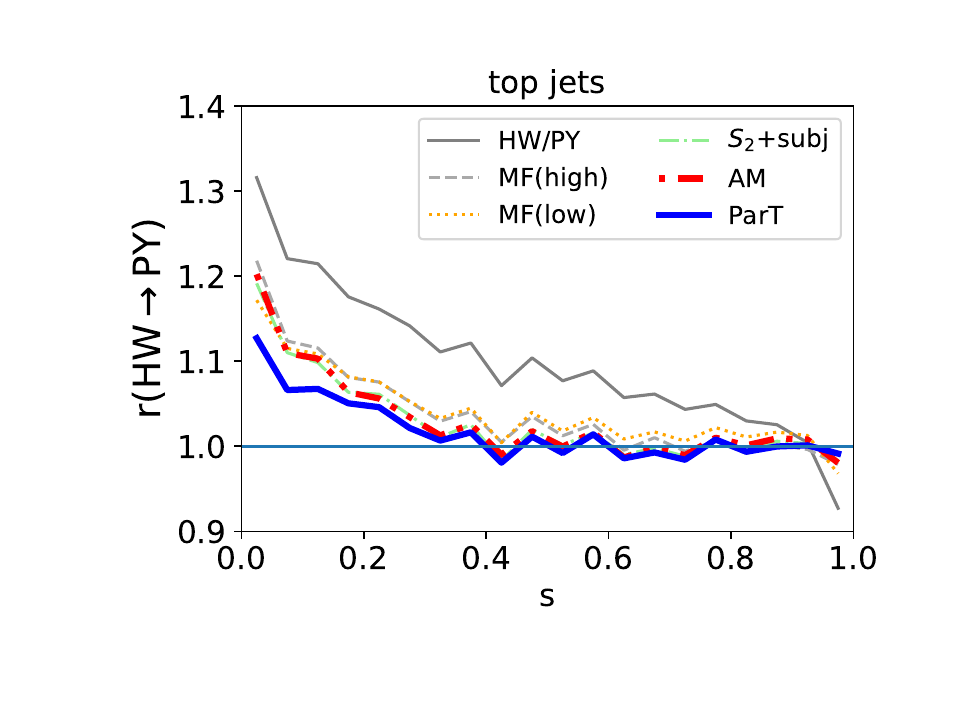}
    \vskip -0.5cm
    \subcaption{ $r( \HW\rightarrow \PY)$ for top jets $\PT$ and $\HT$}
    \end{minipage}
    \caption{
The histogram ratios of the reweighted HW distribution to the PY distribution $r( \HW\rightarrow \PY)=r(s)$ for the ParT output $s$ trained for $\PT$ vs.~$\PQ$.
    The lines represent different reweighting models: MF(high) (grey-dashed), MF(low) (orange-dotted), $S_2+\subj$ (green-dot-dashed), AM (red-dot-dashed), and ParT (blue-solid).   
   The grey lines are the ratio between HW and PY jets without using reweighting.
    }
    \label{fig:re_sratio}
\end{figure}

In \figref{fig:re_sratio}, we show the histogram ratio $r(s)$ for the reweightings using the AM-PIPs.
\figref{fig:re_sratio}(a) is for QCD jets $\PQ$ and $\HQ$ and \figref{fig:re_sratio}(b) is for top jets $\PT$ and $\HT$.
The grey line represents the ratio between HW and PY before the reweighting, i.e., $N(s|\mathrm{HW}) / N(s|\mathrm{PY})$.
The reweighting will bring this grey line closer to $r(s)=1$. 

For the QCD jets, MF(high) shows the worst performance as it considers only hard jet constituents whose systematic uncertainties are less significant compared to soft constituents. MF(low) considers
prominent systematics in soft object modelling, and the ratio becomes closer to one compared to MF(high). $S_2+\subj$ has similar performance with MF(low) and mainly focus on the IRC-safe 
features. $S_2+\subj$ and MF(low) are less correlated in the QCD jet distributions. Therefore, AM performance is much better than any other AM-PIP.  
\begin{table}[ht]
\begin{center}

\begin{tabular}{c|rrrrrr}
\toprule
reweighting model & MF(low)& MF(high)& $S_2+\subj$ & AM & ParT \cr
\midrule
$s<0.05$ & $-0.027$& $-0.067$ & $-0.025$ & $-0.007$ & $0.010$\cr 
$s \in (0.5, 0.7)$& 0.036& 0.100 & 0.027 & 0.003 & $-0.016$\cr
\bottomrule
\end{tabular}
\caption{
Deviations of normalized weight sum ratio in \eqref{eq:rwratio} from one, $r(s)-1$, for reweighted HW QCD jets to PY QCD jets using various reweighting models listed in the first row. 
We consider two ranges: QCD jet-like region with $s<0.05$, and mildly top jet-like region with $s \in (0.5, 0.7)$.
All the results are using rescaled weights.
}\label{tab:qcdratio}
\end{center}
\end{table}

For a quantitative comparison in a specific classifier output range, \tabref{tab:qcdratio} shows the ratio for the QCD jets in QCD-like regions ($s < 0.05$) and moderately top-like regions ($s \in (0.5, 0.7)$). 
The top-like region ($s > 0.7$) is not included due to limited sample numbers.
From the table, we observe that the reweighting quality improves by up to 3\% for MF(low), which is three times better than for MF(high). However, for the best performance, all features need to be considered.

For top-like jets,  $S_2+\subj$ achieves reweighting results compatible with AM, showing the utilization of IRC safe information is dominant in this reweighting for the classification.
\tabref{tab:topratio} presents the ratio for top jets in top-like region ($s > 0.95$) and moderately QCD-like region ($s \in (0.3,0.5)$). In these regions, the deviation is relatively small for ParT as the reweighting works for  $s >0.3$, while AM and $S_2+\subj$ are worse by a factor of two to three.  
The quantitative discussion on the QCD-like top jets will be found in Section 4. 

\begin{table}[ht]
\begin{center}
\begin{tabular}{c|rrrrr}
\toprule
reweighting model & MF(low)& MF(high)& $S_2+\subj$ & AM & ParT \cr 
\midrule
$s>0.95$ & $-0.032$ & $-0.021$ & $-0.017$ & $-0.019$ & $-0.009$ \cr
 $s \in (0.3,0.5)$ & 0.030 & 0.027 & 0.010 & 0.011 & 0.003 \cr
\bottomrule
\end{tabular}
\caption{
Deviations of normalized weight sum ratio in \eqref{eq:rwratio} from one, $r(s)-1$, for reweighted HW top jets to PY top jets using various reweighting models listed in the first row. 
We consider two ranges: top jet-like region with $s>0.95$, and mildly QCD jet-like region with $s \in (0.3, 0.5)$.
All the results are using rescaled weights.
}
\label{tab:topratio}
\end{center}
\end{table}

In summary, we found that the generator classifiers trained by the QCD jet datasets and the top jet datasets focus on different features. 
For the systematics of QCD jets, the difference in the soft activity encoded in MF(low) plays a significant role. 
For the top jets, the difference of the generators is adequately summarized in $S_2+\subj$, and the role of MF and counting variables is subleading. 
We recall the HLFs considered in AM are mainly motivated for top jet tagging, not for reweighting. 
As a consequence, there are still some remaining gaps between ParT reweighting performance in top jets and that of AM (see \tabref{tab:topratio}), although the HLFs are quite robust.
The gap in top jets reweighting between ParT and AM still needs to be understood. 
\secref{sec:results} will delve into the cause of this difference and explain the impact of correlations in collimated high-energy jet components by utilizing energy flow polynomials (EFPs).

\section{Calibrating EFP--Role of independent HLFs}\label{sec:results}
\label{section4}
In this section, we explore the role of Energy Flow Polynomials (EFPs) \cite{Komiske:2017aww, Komiske:2018cqr} in identifying residual discrepancies between PY and HW after the reweighting using AM. 

\subsection{Generator Dependence of EFP Distributions and Event Reweighting }\label{subsec:EFP}

In the previous sections, we demonstrated that AM performs exceptionally well when reweighting HW QCD samples. 
Meanwhile, both AM and ParT models have problems reweighting top samples in the QCD-like regions, and ParT is better than AM in reweighting.
These results indicate that the selected HLF sets are suitable for describing top jet tagging and HW-generated QCD jet reweighting, but additional HLFs are required to describe the top jet reweighting fully.
As the difference between PY and HW appears in details of parton shower and hadronization setups, it is worth considering higher-order correlations.

To address this challenge, we employ Energy Flow Polynomials (EFPs) \cite{Komiske:2017aww}, which have been proposed as promising HLFs \cite{Komiske:2018cqr}.
EFPs extract the energy distribution and spatial correlations among jet constituents in a permutation-invariant manner, offering potential improvements in classifier performance. 
By analyzing the discrepancies as functions of EFPs, we will attempt to identify the feature spaces with large deviation that AM cannot reweight. 

EFPs are defined using a multigraph $G$ as follows:
\begin{align}
\mathrm{EFP}_G = \sum_{i_1=1}^M \dots \sum_{i_N=1}^M z_{i_1}^{\kappa} \times \cdots \times z_{i_N}^{\kappa} \prod_{(k,l)\in G} \theta_{i_k i_l}^\beta,
\end{align}
where:  
\begin{itemize}
    \item $z_i = p_{\mathrm{T},i}/p_{\mathrm{T},\mathrmbf{J}}$ represents the momentum fraction of the constituent $i$,
    \item $\theta_{ij} = \sqrt{\Delta y_{ij}^2 + \Delta \phi_{ij}^2}$ denotes the spatial distance between constituents $i$ and $j$,
    \item $M$: number of jet constituents.
\end{itemize}
EFPs reflect various aspects of jet substructures depending on the configuration of the graph $G$ and the parameters $\kappa$ and $\beta$.
Note that for $\kappa\neq 1 $, these EFPs cover energy-dependent IRC-unsafe features, providing an additional set of variables that is not  covered by  IRC-safe energy correlators, and MFs that are geometric features and energy-independent. 

\begin{table}[t!]
    \centering
\begin{tabular}{c|cc|l|c}
    \toprule
  $n$ & $\kappa$ & $\beta$ & graph $G$ & graph diagram
   \\
   \midrule
    $\FD_1$ & 2& 2 & $(0,1), (0,2), (1,2)$&
\begin{tikzpicture}[baseline={([yshift=-0.8em]current bounding box.north)}]
\draw[-] (0,0) -- ++(60:0.5) -- ++(180:0.5) -- cycle;
\node[anchor=north] (bottom) at (0,-0.1) {};
\end{tikzpicture} 
\\ 
    $\FD_2$ & 2& 1 &$ (0,1), (0,2), (1,2)^2, (2,3)^3$&
\begin{tikzpicture}[baseline={([yshift=-0.8em]current bounding box.north)}]
\draw[-] (0,0) -- ++(60:0.5) -- ++(180:0.5) -- cycle;
\draw[-] (0,0)++(60:0.5) 
    to[out=335,in=205,distance=-0.2cm]
    ++(180:0.5);
\draw[-] (0,0)++(60:0.5) 
    --
    ++(30:0.5);
\draw[-] (0,0)++(60:0.5) 
    to[out=185,in=55,distance=-0.2cm]
    ++(30:0.5);
\draw[-] (0,0)++(60:0.5) 
    to[out=235,in=5,distance=-0.2cm]
    ++(30:0.5);
\node[anchor=north] (bottom) at (0,-0.1) {};
\end{tikzpicture}   
    \\ 
    $\FD_3$ & 0& 1 & $(0,1), (0,2), (0,3), (0,4), (0,5), (0,6), (0,7), (0,8)$&
\begin{tikzpicture}[baseline={([yshift=-0.8em]current bounding box.north)}]
\draw[-] (0,0) -- ++(0:0.5);
\draw[-] (0,0) -- ++(45:0.5);
\draw[-] (0,0) -- ++(90:0.5);
\draw[-] (0,0) -- ++(135:0.5);
\draw[-] (0,0) -- ++(180:0.5);
\draw[-] (0,0) -- ++(225:0.5);
\draw[-] (0,0) -- ++(270:0.5);
\draw[-] (0,0) -- ++(315:0.5);
\node[anchor=north] (bottom) at (0,-0.6) {};
\end{tikzpicture}
    \\
    $\FD_4$ & 1 & 0.5  & $(0,1)^2,  (0,2)^2, (1,2)^3$ &
\begin{tikzpicture}[baseline={([yshift=-0.8em]current bounding box.north)}]
\draw[-] (0,0) -- ++(60:0.5) -- ++(180:0.5) -- cycle;
\draw[-] (0,0) 
    to[out=210,in=85,distance=-0.2cm] 
    ++(60:0.5) 
    to[out=335,in=205,distance=-0.2cm]
    ++(180:0.5) 
    to[out=95,in=325,distance=-0.2cm]
    cycle;
\draw[-] (0,0)++(60:0.5) 
    to[out=25,in=155,distance=-0.2cm]
    ++(180:0.5);
\node[anchor=north] (bottom) at (0,-0.1) {};
\end{tikzpicture}    
    \\ 
    $\FD_5$ & 1 & 1 & $(0,1), (0,2), (1,2), (3,4), (3,5), (4,5), (2,3)$ &
\begin{tikzpicture}[baseline={([yshift=-0.8em]current bounding box.north)}]
\draw[-] (0,0) -- ++(30:0.5) -- ++(150:0.5) -- cycle;  
\draw[-] (0,0)++(30:0.5) -- ++(0:0.5) -- ++(330:0.5) -- ++(90:0.5) -- ++(210:0.5);
\node[anchor=north] (bottom) at (0,-0.1) {};
\end{tikzpicture}
    \\
    $\FD_6$ &  2 & 0.5 & $(0,1), (0,2), (1,2)^2, (0,3), (2,3)$&
\begin{tikzpicture}[baseline={([yshift=-0.8em]current bounding box.north)}]
\draw[-] (0,0) -- ++(0:0.5) -- ++(120:0.5) -- cycle;  
\draw[-] (0,0)++(0:0.5) -- ++(60:0.5) -- ++(180:0.5);  
\draw[-] (0,0)
    to[out=155,in=25,distance=-0.2cm]
++(0:0.5);  
\node[anchor=north] (bottom) at (0,-0.05) {};
\end{tikzpicture}
    \\ 
    \bottomrule
    \end{tabular}
    \caption{The first six features of $\FD_G\coloneqq\log({\rm EFP}_G )$ contributing to  top vs.~QCD classification for the PY datasets selected by DisCo \cite{Das:2022cjl}. The second column from the last (graph $G$) indicates the connected vertices ($=\text{constituents}$) $i$ and $j$ as $(i,j)$ format, and the last column shows its diagrammatical representations
    . The column $\kappa$ indicates the power of momentum, and $\beta$ is the power of spatial distance. See text. 
    }
    \label{tab:FD}
\end{table}

In \cite{Das:2022cjl}, it has been shown that the Distance Correlation (DisCo) method is an efficient approach to selecting important features that contribute significantly to top vs.~QCD classification performance from a large pool of EFP candidates.  
In the study, the six key EFPs in \tabref{tab:FD} are identified from over 7,000 candidates based on their ability to optimize the classifier's $R_{30\%}$ for top vs.~QCD classification.\footnote{The EFPs were selected after starting from the initial three features $m_\mathrmbf{J}, p_{\mathrm{T}}$ and $m_W$, where  $m_W$ is $W$ boson candidate mass. Therefore, EFPs of two vertices do not show up in the key EFPs of \figref{tab:FD}.}

In this section, we denote the leading six EFPs as $\FD_n\coloneqq\log(\mathrm{EFP}_n)$ $(n=1,\cdots,6)$. 
The role of these EFPs may be understood as follows;
$\FD_1$ and $\FD_4$ capture the three-pronged structure of top jets, while $\FD_3$, with $\kappa=0$, highlights contributions from soft particles.  
$\FD_1$, $\FD_2$, and $\FD_6$, characterized by $\kappa=2$, are sensitive to high-energy particles.  
Finally, $\FD_2$ and $\FD_6$ involve more than four vertices
and capture the internal substructure of the three leading clusters of the top decay.

\begin{figure*}[ht!]
\centering
\includegraphics[width=0.4\linewidth]{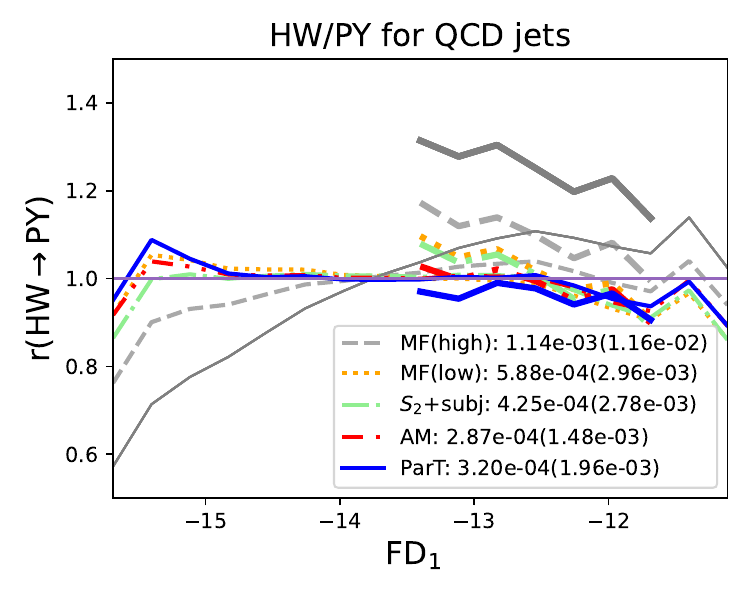}
\includegraphics[width=0.4\linewidth]{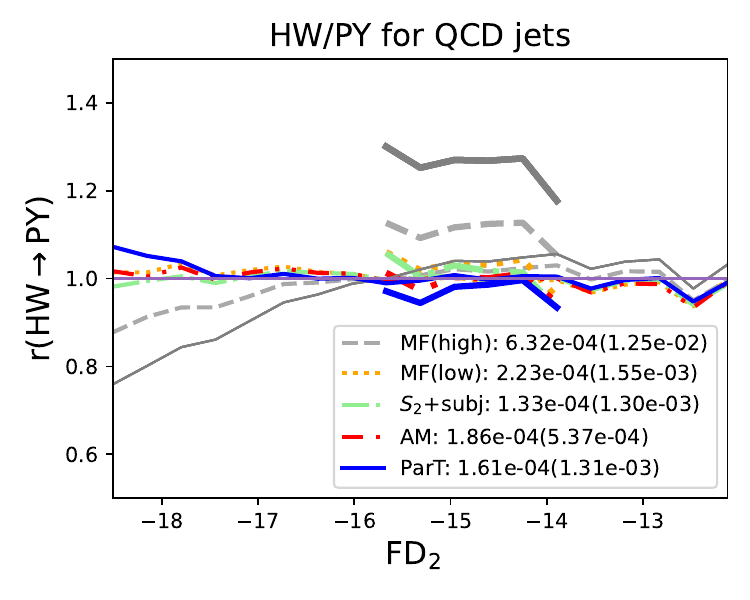}
\\
\vspace{0.2cm}
\includegraphics[width=0.4\linewidth]{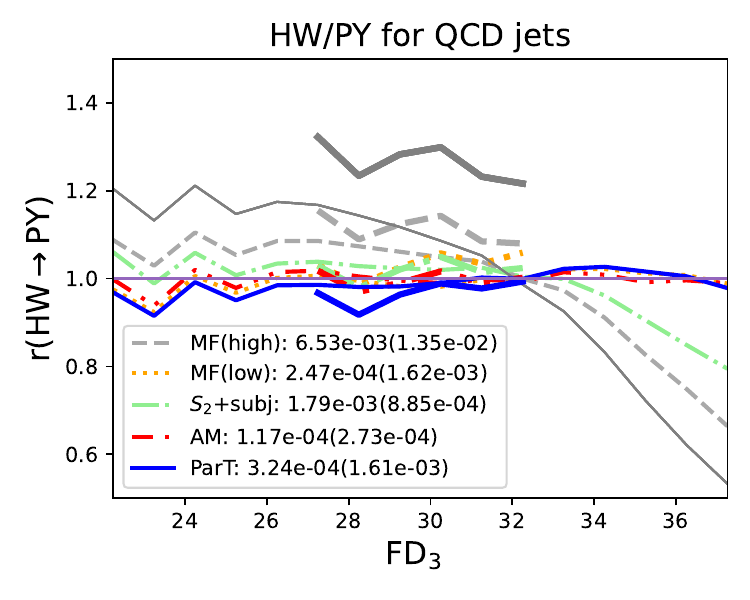}
\vspace{0.2cm}
\includegraphics[width=0.4\linewidth]{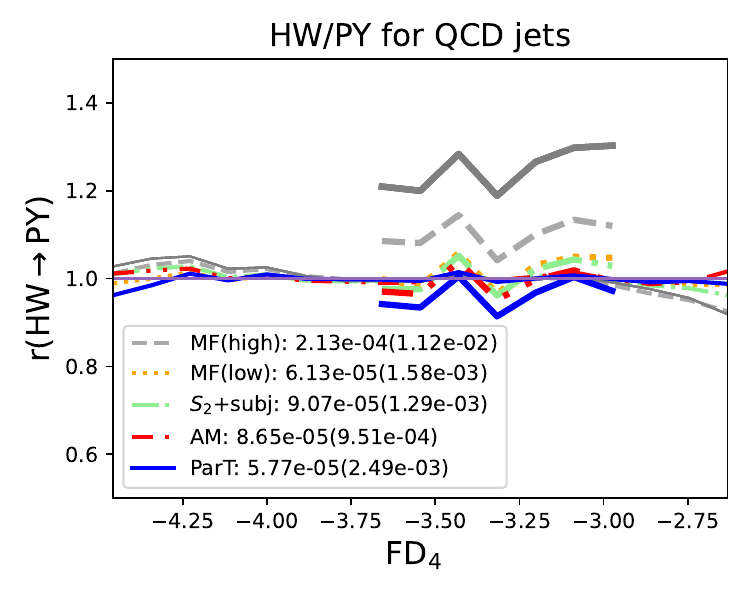}
\\
\vspace{0.2cm}
\includegraphics[width=0.4\linewidth]{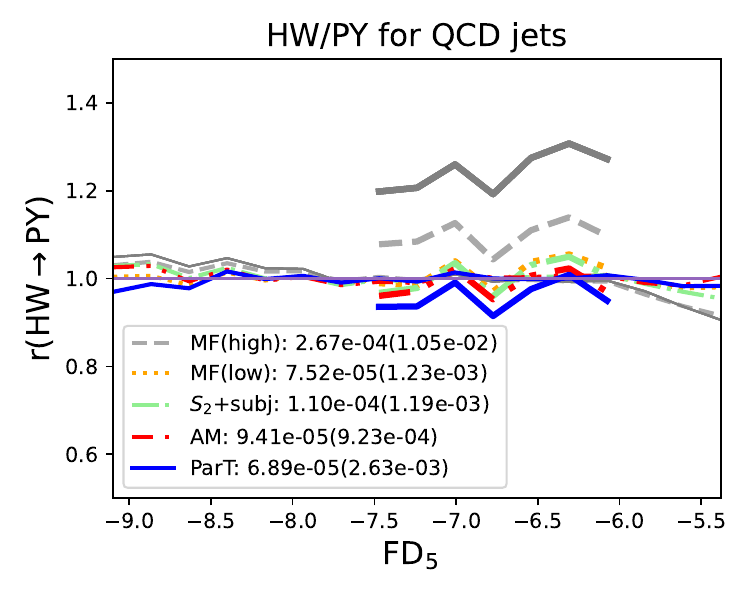}
\includegraphics[width=0.4\linewidth]{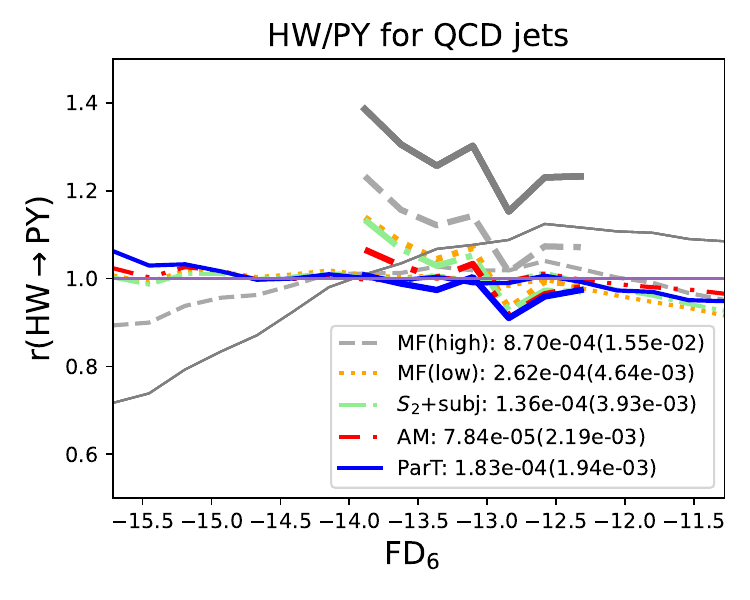}
\caption{
The histogram ratios of the reweighted HW distribution to the PY distribution $r(\HW\rightarrow \PY) =r(\FD_n)$
for all QCD jets (thin lines) and the QCD jets after the cut $s>0.8$ (the thick lines). 
The reweighting accuracy $\Delta$  is displayed after each legend. The numbers inside brackets are for the jets with $s>0.8$. The lines correspond to the different reweighting models;  MF(high) (light grey dashed), $\MF$(low) (orange dotted),  $S_2+\subj$ (light green dot-dashed), AM (red dot-dot-dashed) and ParT (blue solid). 
The grey lines show the ratio before the reweighting. If the corresponding reweighting model accurately calibrates the HW dataset into the PY dataset, the ratio will be close to one. 
We only plot the ratio with more than 1,000 events in the bin for both $\HW$ and $\PY$ jets.}
\label{fig:EFP_pyhw}
\end{figure*}

\begin{figure}[ht!]
\centering
\includegraphics[width=0.4\linewidth]{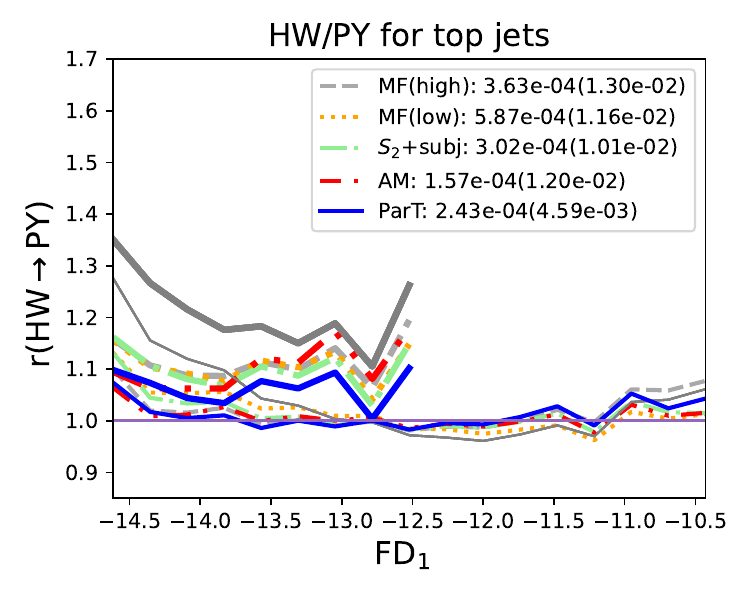}
\includegraphics[width=0.4\linewidth]{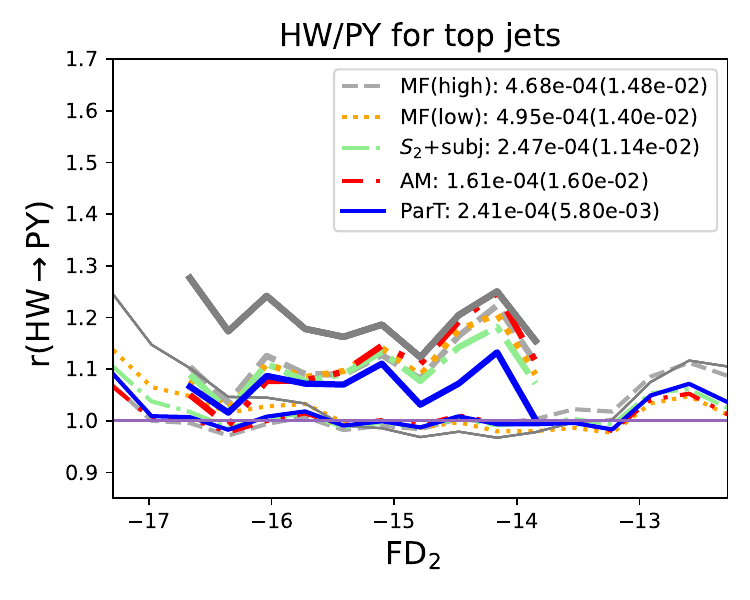}
\\
\vspace{0.2cm}
\includegraphics[width=0.4\linewidth]{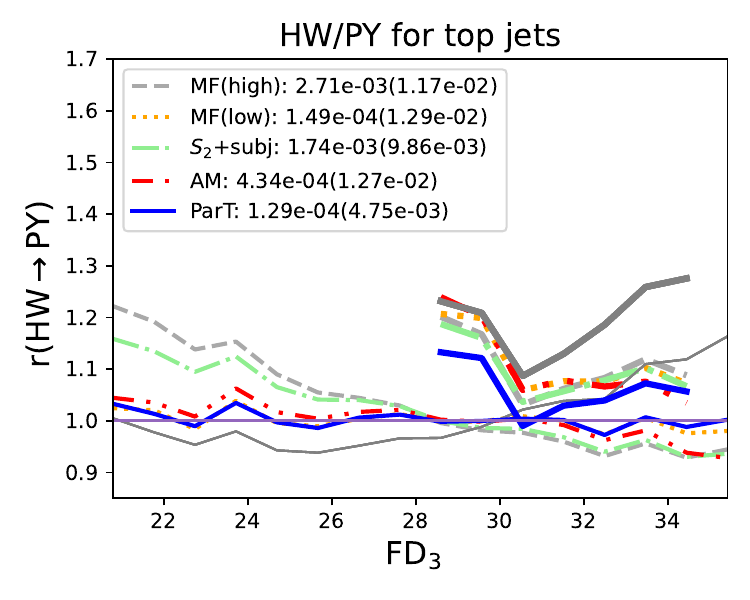}
\includegraphics[width=0.4\linewidth]{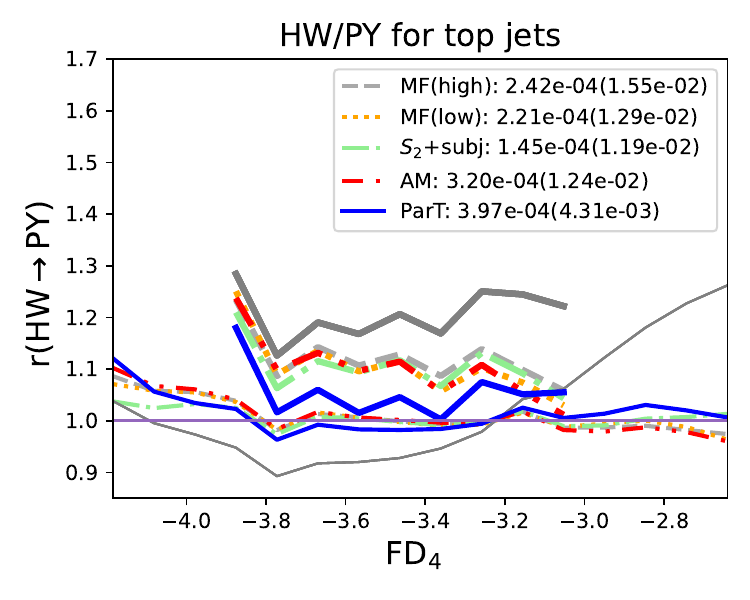}
\\
\vspace{0.2cm}
\includegraphics[width=0.4\linewidth]{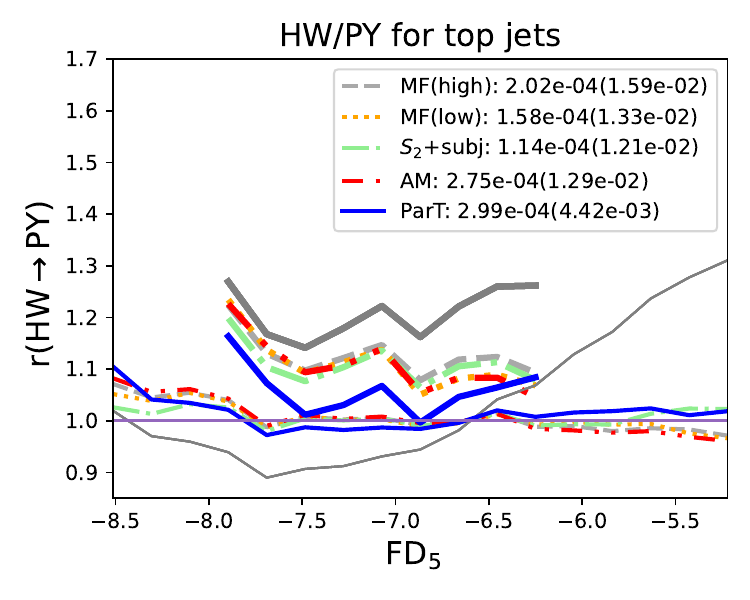}
\includegraphics[width=0.4\linewidth]{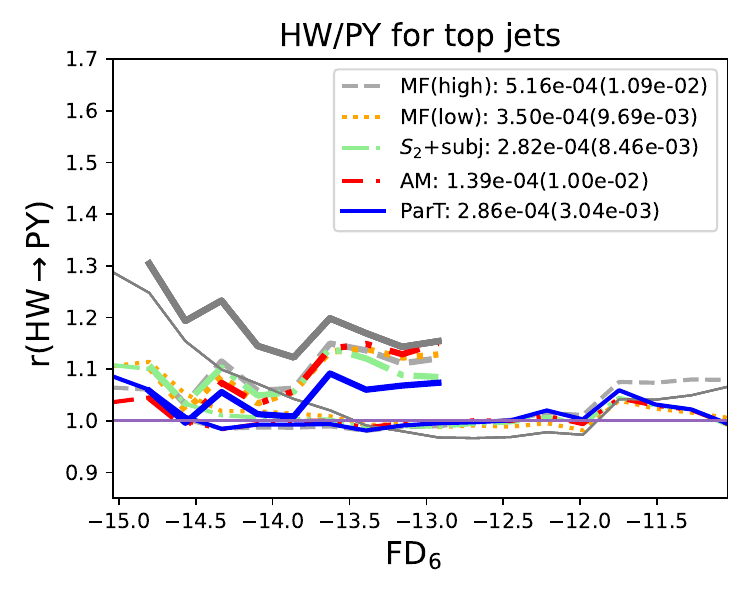}
\caption{
The histogram ratios of the reweighted HW distribution to the PY distribution for the top jets $r(\HW\rightarrow \PY)=r(\FD_n)$ for all top jets  (thin lines) and top jets after the cut $s<0.2$ (QCD-like top jets, thick lines). 
The line styles are the same as \figref{fig:EFP_pyhw}.  
The reweighting accuracy $\Delta$ is displayed after each legend, and the numbers in the bracket 
is for the QCD-like top jets.
The matching to the $\PT$ distribution is poor for the QCD-like top jets, especially at the higher $\FD_1$, $\FD_2$ and $\FD_6$ and the smaller $\FD_3$, $\FD_4$ and $\FD_5$. 
}
\label{fig:EFP_pyhwtop}
\end{figure}

In the following, we present the reweighting analysis of the EFP distributions,  comparing the performance of the ParT, AM, and AM-PIPs. 
The EFP values are obtained by using the package \cite{Komiske:2017aww, energyflow}. AM 
relies on preselected HLFs; therefore, it might result in poor reweighting performance in some regions of the parameter space if HLFs of the AM lack specific information encapsulated by $\FD_n$. 
On the other hand, ParT directly extracts features relevant to classification from LLFs, and the chance of such bias occurring is smaller.

We show the ratio of histograms 
\begin{equation}
r(\FD_n)= \frac{N(\FD_n \vert \HW; w)}{ N(\FD_n\vert \PY)},
\end{equation}
where $N(\FD_n|\HW;w)$ is the weight sum fraction of HW jets after reweighting in a $\FD_n$ bin, and $N(\FD_n|\PY)$ is the fraction of bare PY jets in the same bin.
To assess the accuracy of the correction in specific regions, we also show top-like QCD jets ($s > 0.8$) and QCD-like top jets ($s< 0.2$).  We have already seen that  ParT and AM reweighting performance is poor for QCD-like top jets. 
To quantitatively evaluate the accuracy of the corrections, we set the number of the HW and PY events to be equal before the reweighting and calculated  $\Delta(\FD_n)$ as follows:  
\begin{align}
\Delta(\FD_n) \coloneqq 
\frac{1}{\sum_i N_i(\FD_n|\PY)} 
\sum_j 
\frac{(N_j(\FD_n|\HW;w) - N_j(\FD_n|\PY))^2}{N_j(\FD_n|\PY)},  \label{eq: delta_fd}
\end{align}
where  $i$ and $j$ is the bin index. 

\figref{fig:EFP_pyhw} shows the results of the reweighting $\FD_n$ distributions of QCD jets for the entire dataset (thin lines) and the top-like jets ($s> 0.8$, thick lines).
Values of the accuracy metric $\Delta(\FD_n)$ are listed in the legend of the lines. The accuracy metric of the top-like jets $s$ is inside the parentheses. Before reweighting, the ratios of $\FD_n$ distributions deviate significantly from unity, except for $\FD_4$ and $\FD_5$, highlighting substantial generator dependence.  
AM consistently outperforms ParT in the accuracy metric for the reweighting of QCD jets. 
The comparison among performance using AM-PIP further identifies the correlation between HLFs of AM and EFP.  
For example, the IRC-unsafe $\FD_3$, $\MF$(low) achieves $\Delta(\FD_3) \sim 2.47 \times 10^{-4}$, while $S_2+\subj$ yields $\Delta(\FD_3) = 1.79 \times 10^{-3}$. 
This relation flips in the top-like region, $S_2+\subj$ achieve better alignments for  with $\Delta(\FD_3) = 8.85 \times 10^{-4}$, compared with $1.62 \times 10^{-3}$ of  $\MF$(low).  This suggests the importance of jet substructure in top-like QCD jets.  

\figref{fig:EFP_pyhwtop} presents the reweighting results of the top jets (thin lines) and the subset of QCD-like top jets ($s < 0.2$, thick lines).
AM and ParT models exhibit similar accuracy for the full dataset, $\Delta(\FD_n)<10^{-4}$. 
For the QCD-like top jets, AM and AM-PIP perform worse than ParT. 
For example, $\Delta(\FD_2) = 1.6 \times 10^{-2}$ for  AM  and $5.8 \times 10^{-3}$ for ParT; the gap between ParT and AM is bigger in large $\FD_2$ region, and AM is worst among all AM-PIP, even though AM have all HLF information.  This suggests that the dominant difference arises in the region blinded from the HLFs of AM. 
AM only fails in the phase space that cannot be described by HLFs, while  AM-PIP fails in wider region 
so that the performance in the large $\FD_2$ region is accidentally better than AM. 
The similar behaviour is observed for $\FD_1$ and $\FD_6$; they are also EFP with $\kappa=2$. 
These findings also suggest the merit of ParT using LLFs directly, even though ParT suffers large statistical fluctuation. The AM HLFs are selected by human prejudice and can fail drastically in some phase space.
The result also indicates the merit of testing the reweighting model by HLFs that are not used for the classifier inputs.

\subsection{Tracking the Discrepancy of ParT and AM in QCD-like top jets}

In this subsection, we focus on the QCD-like top jet and qualitatively identify the phase space that is not described by HLFs of AM but contributes to the generator classification (PY vs.~HW). 

In \figref{fig:EFP_pyhwtop}, the reweighting accuracy of AM depends significantly on the $\FD_n$ values for QCD-like top events with $s<0.2$: 
the reweighting accuracy is worse in regions with high (low) $\FD_n$ values for $\FD_1$, $\FD_2$, and $\FD_6$ ($\FD_3$, $\FD_4$, and $\FD_5$), respectively. 

\begin{figure}[ht!]
    \centering
 \begin{minipage}[t]{0.45\linewidth}
    \includegraphics[width=0.9\linewidth]{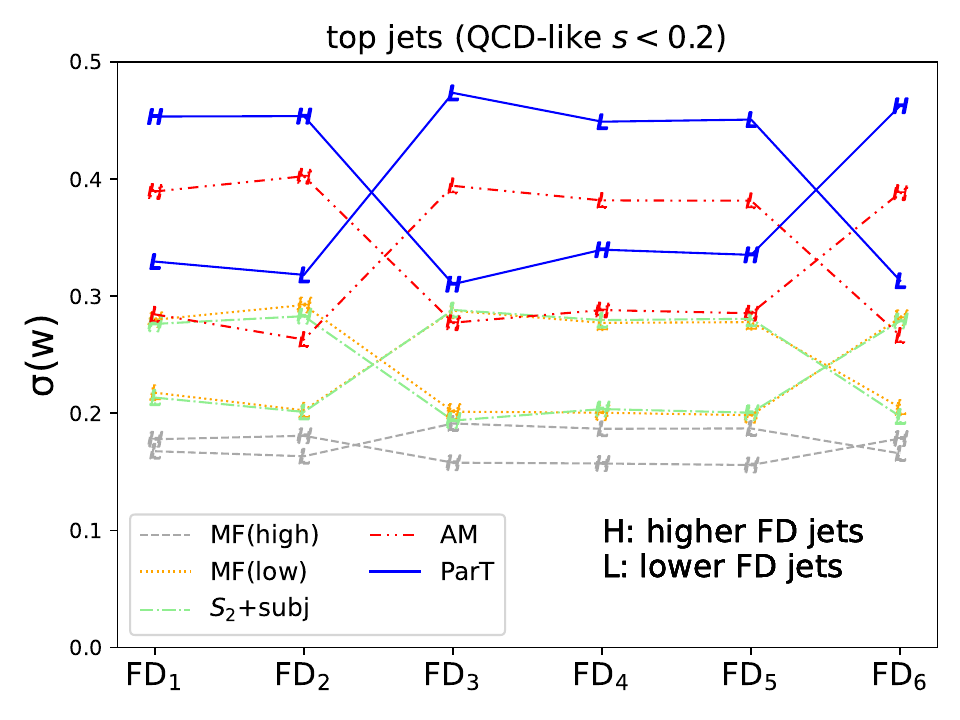}
    \subcaption{The $\sigma(w)$ of QCD-like top event ($s <0.2$)}
    \end{minipage}
    \hskip 0.04\linewidth
     \begin{minipage}[t]{0.45\linewidth}
   \includegraphics[width=0.9\linewidth]{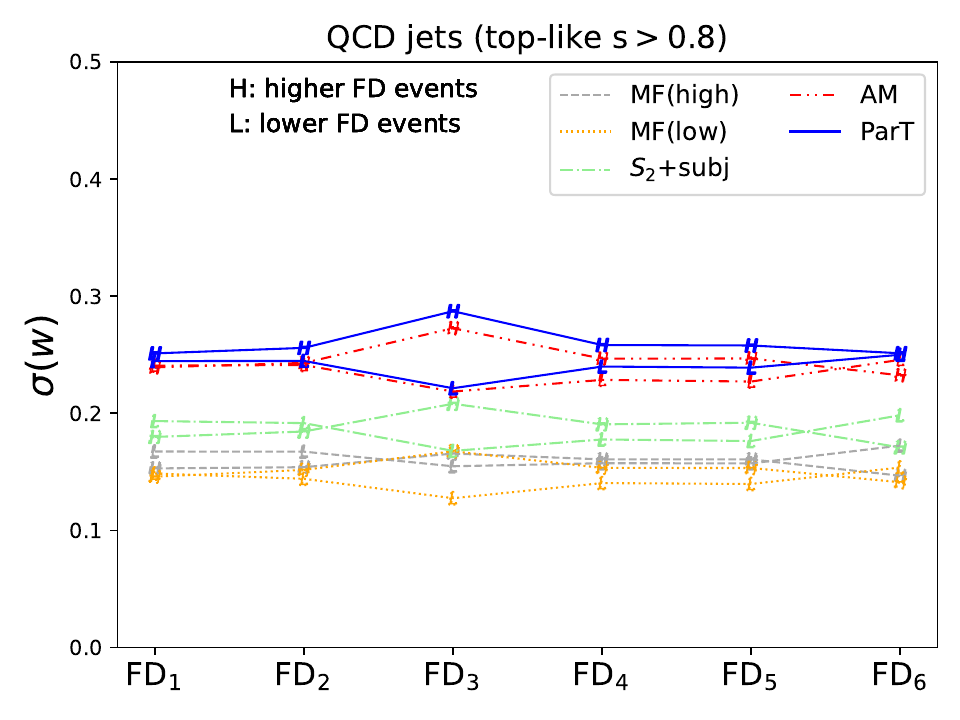}
   \subcaption{The $\sigma(w)$ of top-like QCD event ($s^ >0.8$)}
   \end{minipage}
    \caption{The standard deviation (std) of the event weight  $\sigma(w)$ ($y$-axis). 
    The points with the label H (L) correspond to the std values of the events in the H-region (L-region) of corresponding $\FD_n$ (the $x$-axis), respectively. 
    The data of the same reweighting models and the same H-region (L-region) are connected for readability, with the colour codes and the line types being the same as \figref{fig:EFP_pyhw}.
    Figure (a) is for QCD-like top jets ($s<0.2$), and  (b) is for top-like QCD jets ($s>0.8$). 
    }
    \label{fig:FDstd}
\end{figure}

The poor accuracy is partially due to the small overlap in the phase space distribution. In such cases, significant events with either very large or small weights lead to large statistical fluctuation.   To show this, we compare the weight distribution in three phase-space regions, low (L), middle (M), and high (H) in $\FD_n$ values, where the cut values are adjusted so that the number of events in each region is equal. 
\figref{fig:FDstd}(a) shows the standard deviation of the weights $\sigma(w)$ of the events in H (high) and L (low)  $\FD_n$ region for QCD-like top events ($s < 0.2$). 
Here, we took models $\{\ParT, \AM, \MF(\low), \MF(\high), S_2+\subj\}$, and the results of the same model and same H (L) regions are connected by a line for visibility.  
A clear hierarchy in $\sigma(w)$ is observed in the plot:
\begin{equation}
    \sigma(w^{\ParT}) > \sigma(w^{\AM}) > \sigma(w^{\MF(\low)}) \sim \sigma(w^{S_2+\subj}) > \sigma(w^{\MF(\high)}).
\end{equation}
Here, the superscript of $w$ denotes the reweighting model considered.
ParT (the blue solid line) exhibited the largest standard deviation for all $\FD_n$, reflecting its ability to cover a broader phase space, where the higher value is above 0.4 for all $\FD_n$. This deviation is large, given the average weight value $\sim 1$. 
Moreover, the values of $\sigma(w^{\ParT})$ in L- and H-regions differ by more than 0.1. 
The region with higher $\sigma(w^{\ParT})$ involves the region with a large PY vs.~HW probability ratio. Therefore, finding the correct reweighting factor requires larger training samples. 
In \figref{fig:FDstd}(b), the same plot for the top-like QCD events ($s > 0.8$) is shown.  The $\sigma(w^{\ParT})$ is typically $\sim 0.2$, and differences of the $\sigma(w)$ values are less pronounced; The values of AM and ParT are very close to each other. Moreover, $\sigma(w)$ values for L- and H-regions are the same except $\FD_3$.  

\begin{figure}[ht]
    \centering
    \begin{minipage}[t]{0.45\linewidth}
\includegraphics[width=0.9\linewidth]{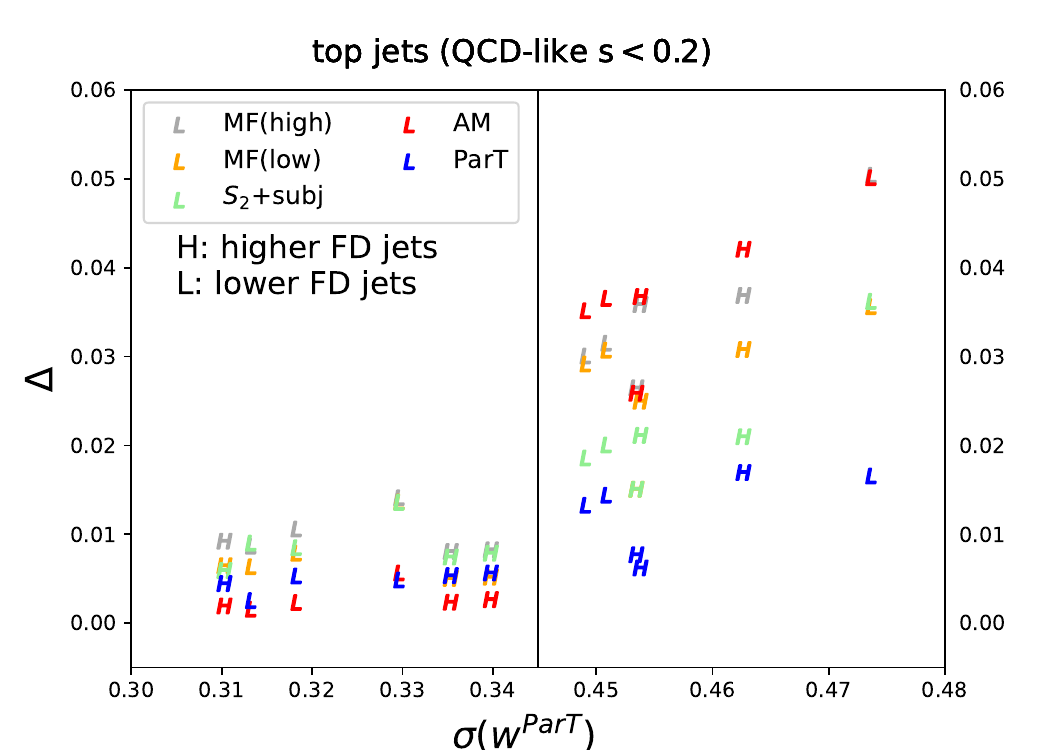} 
     \subcaption{The plots of $\Delta$ vs.~$\sigma(w^{\ParT})$ of QCD-like top events.}
    \end{minipage}
    \hskip 0.04\linewidth 
    \begin{minipage}[t]{0.45\linewidth}
\includegraphics[width=0.9\linewidth]{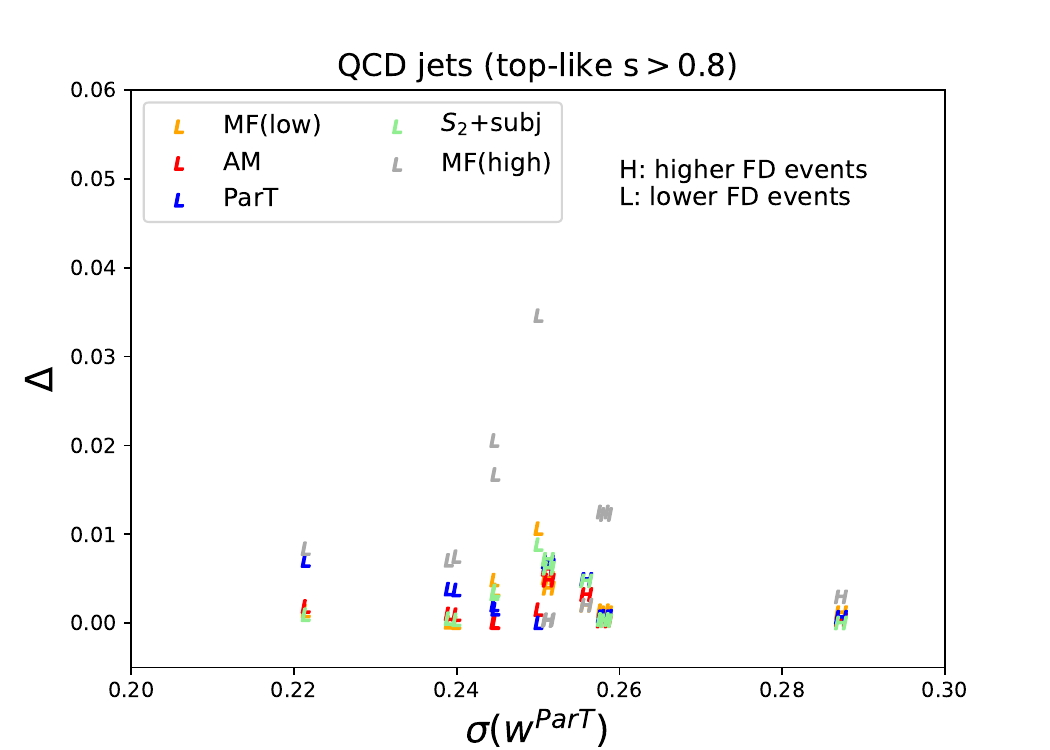}   
       \subcaption{The plots of $\Delta$ vs.~$\sigma(w^{\ParT})$ of top-like QCD events.}
    \end{minipage}

    \caption{The correlation between the reweighting accuracy $\Delta_\mathrm{L(H)}$ and $\sigma(w^{\ParT})$. The $y$-axis is  $\Delta_\mathrm{L(H)}$ of the jets in the H-region (L-region) of $\FD_n$, and $x$-axis is $\sigma(w^\ParT)$ of the corresponding dataset. The colour code of "H" and "L" marks is the same as \figref{fig:EFP_pyhw}. 
    Figure (a) is for QCD-like top jets ($s<0.2$), and 
     (b) is for top-like QCD jets ($s>0.8$).    
    }
    \label{fig:Rvswstd}
\end{figure}

The standard deviation $\sigma(w^{\ParT})$ tends to be large in the region where AM reweighting performance is poor. To quantify this, we plot $\sigma(w^{\ParT})$ vs.~$\Delta_\mathrm{H(L)}(\FD_n)$ in \figref{fig:Rvswstd}. Here accuracy metric $\Delta(\FD_n)_\mathrm{H(L)}$ is defined as follows, 
\begin{equation}
\Delta_\mathrm{L}(\FD_n) \coloneqq \left( \frac{N_\mathrm{L}(\FD_n|\HW;w)}{N_\mathrm{L}(\FD_n|\PY)} - 1 \right)^2, \quad
\Delta_\mathrm{H}(\FD_n) \coloneqq \left( \frac{N_\mathrm{H}(\FD_n|\HW;w)}{N_\mathrm{H}(\FD_n|\PY)} - 1 \right)^2,
\end{equation}
where $N_\mathrm{L}$ and $N_\mathrm{H}$ represent the number of (reweighted) events in the L- and H-regions.

\figref{fig:Rvswstd}(a) illustrates the relationship between $\sigma(w^{\ParT})$ and $\Delta(\FD_n)$ for QCD-like top events. In regions with high $\sigma(w^{\ParT})$, namely, H-regions of $\FD_1$, $\FD_2$, and $\FD_6$, and L-regions of $\FD_3$, $\FD_4$, and $\FD_5$, the AM  values (marked by red "H" and "L")  of $\Delta_\mathrm{H(L)}(\FD_n)$ are the largest, namely the worst in accuracy.  
Conversely, in the left panel of \figref{fig:Rvswstd}(a), namely, the regions with lower $\sigma(w^{\ParT})$, AM has the best reweighting accuracy. \figref{fig:Rvswstd}(b) show the $\sigma(w^{\ParT})$ and $\Delta(\FD_n)$ for top-like QCD jets, where $\sigma(w^{\ParT})$ does not correlated with $\Delta$

The relations observed in \figref{fig:FDstd}(a), $\sigma(w^{\ParT}) \gg \sigma(w^{\AM})$, and the right panel of \figref{fig:Rvswstd}(a), $\Delta^\AM(\FD_n)\gg  \Delta^\ParT(\FD_n)$ for large $\sigma^{\ParT}$ and $\Delta^\AM(\FD_n)\ll \Delta^\ParT(\FD_n)$ for small  $\sigma^{\ParT}$, indicate the lack of appropriate HLFs representing the difference between $\HT$ and $\PT$ samples in high $\sigma(w^{\ParT})$ region of $\FD_n$. 
The reduced reweighting accuracy of AM can be attributed to the following factors:
\begin{itemize}
    \item  Jet geometry: The discrepancy between ParT and AM occurs in the QCD-like top region, where jets tend to be more collimated.
    It is difficult to see the typical three-prong structure of the top jets in this region. 
    \item The  $\FD_1$, $\FD_2$, and $\FD_6$ are EFPs of $\kappa=2$. 
    In the H-region, a few high-energy jet constituents carry most of the jet energy to get an amplification of $z_i^2$. 
    \item In the L-region of $\FD_3$, the jet must be narrow  to get the suppression by   $\theta_{ij}$.  
\end{itemize}
Namely, AM has a problem with highly collimated top jets containing high-energy jet constituents.  
The AM HLFs do not cover the parameter region described above. For MF calculation, the lower energy cuts of the jet constituents run up to $p_{\mathrm{T}}>8$~GeV. Therefore, the correlations among the jet constituents of $p_{\mathrm{T}}\gg$ 8~GeV are diluted. The subjet module inputs are the subjets momentum of the cone sizes  $R=0.1, 0.2$, and $0.3$; therefore, correlation within $R<0.1$ is masked in the input. 

In summary, it is reasonable that AM performs inferiorly in the regions described above. Indeed, the AM reweighting behaves as if it does not know much about the PY and HW differences; the corrected ratio of AM is close to the original PY vs.~HW ratio in \figref{fig:EFP_pyhwtop}. 
Finally, incorporating $\FD_n$ into the AM model could improve the accuracy of the AM in this region, although extensions of AM are beyond the goal of this study. First, the small overlap between the PY and HW datasets in the phase space indicates more data needed for the training, but top contamination into the QCD region is phenomenologically uninteresting. Moreover, 
The discrepancy of the correlations among high $p_{\mathrm{T}}$ jet constituents should be improved by the advancement of parton shower modelling, 
rather than post-generation reweighting. 

\section{Summary and Discussion}
\label{section5}
Machine Learning (ML) is a powerful tool in particle physics. 
One of the well-established applications of Machine Learning is the reweighting of simulated events \cite{Sugi:2012, Cranmer:2015bka, Brehmer:2018eca, Nachman:2020fff}. 
This is because the event classification task is equivalent to the probability ratio estimate between the target distribution and the simulated distributions. 
In the context of jet physics, the predictions of the event distributions of commonly used event generators such as \texttt{Pythia}, \texttt{Herwig} and \texttt{Sherpa} still show non-negligible deviations from the observed distributions.\footnote{See http://mcplots.cern.ch for comparisons \cite{Karneyeu:2013aha}. } 
Therefore, it is necessary to apply post-hoc reweighting of the events to reduce the uncertainty of MC extrapolations.
Furthermore, ML classifiers can pinpoint regions in phase space where theoretical predictions require refinement, thereby aiding the development of more accurate event simulations.

Modern deep learning models, including graph neural networks \cite{Qu:2019gqs} and transformers \cite{NIPS2017_3f5ee243, shleifer2021normformer, Touvron_2021_ICCV, Qu:2022mxj}, leverage low-level features (LLFs) to achieve high classification accuracy. While utilizing LLFs allows deep learning models to identify complex correlations without human intervention, their lack of interpretability limits their potential to improve our understanding of the simulations. 
We address this problem by using Analysis Model (AM) \cite{Furuichi:2023vdx} 
utilizing the high-level features (HLFs), consisting of IRC-safe and IRC-unsafe quantities. The HLFs we have chosen 
are motivated by QCD and the physics of top jets, have clear geometrical interpretation, and are less sensitive to experimental noise. Therefore, using AM enhances the interpretability of the reweighting process by analyzing HLFs.
AM is a modular network that assesses the individual contribution of HLF, providing insights that can guide the tuning of event generators. By pointing to the key features for model tuning, AM can improve the general agreement between the real data and simulated data.

In this paper, we test the idea of interpretable reweighting for top vs.~QCD classification, setting the Pythia-generated events as the target distribution and reweighting Herwig events by the weights obtained from classifiers identifying event generators. We disentangle the key difference of generator prediction in terms of the HLFs of AM, and compare the AM results with more advanced Particle Transformer  (ParT) to check the equivalence.   
Our study demonstrated that AM reweights QCD jet distributions more precisely than ParT, even in regions contaminating the top-like signal region. This advantage arises because AM consists of only essential permutation invariant HLFs describing jet substructure. As AM is a simpler model than ParT but with similar performance, this property allows efficient training and reduces statistical uncertainties.

However, AM and ParT encounter challenges for QCD-like top events. The difficulties stem from the limited overlap between the PY and HW distributions in some areas of phase space, requiring larger training datasets for effective reweighting. Further investigation using Energy Flow Polynomials (EFPs) \cite{Komiske:2017aww, Komiske:2018cqr} revealed that AM's performance is reduced in the phase space, which is not covered by the selected HLFs, that is, the correlation among the collimated high $p_T$ constituents. 
In the other region, AM retains superior reweighting accuracy over ParT.  Incorporating additional features like EFPs could improve AM's accuracy further, although this is outside the scope of the current study. ParT, by contrast, exhibits more uniform performance across EFP-based reweighting tasks. ParT has higher representation power and identifies discrepancies to target distributions without introducing significant bias, even though it suffers more statistical fluctuation than AM.

In summary, AM using HLFs models studied in this paper, offers a robust and interpretable approach to event reweighting and jet classification. AM enables a systematic comparison of event generators by providing a deeper understanding of ML model performance, and this may be used to reduce systematic uncertainties in collider analyses. Future work could focus on expanding HLFs to improve the power to distinguish the event generators, assessing the reliability of improved event generators, and refining event generators for a better description of 
high-energy collider data, ultimately advancing the precision of theoretical and experimental studies in particle physics.

Finally, in the era of deep learning, it is highly useful to have accurate parton showers and hadronization models that describe jet phenomenology in a wide range of angular and momentum scales. 
This is because deep learning automatically extracts information from low $p_T$ particles for jet identification \cite{Komiske:2018cqr, deOliveira:2015xxd, Lim:2020igi, Dreyer:2020brq}.  
Correlations among jet constituents are sensitive to parton shower and hadronization modelling.  The source of the current disagreement between the simulations and data may come from the insufficient QCD approximation.  It is expected that accurate formulations of parton showers may improve the difference between experimental data and event simulations in the near future \cite{vanBeekveld:2023ivn, Karlberg:2021kwr, Dasgupta:2018nvj}. Nevertheless, the prediction of the soft or collimated particle distributions is limited by our understanding of strong dynamics in parton shower and hadronization processes, and some post-generation refinement would still be useful.  
AMs using HLFs would provide solid ground for comparing simulation and data and improving the agreement between them, making them ready for deep learning analysis in HL-LHC.

\acknowledgments

This work is supported by Grant-in-Aid for Transformative Research Area (A) 22H05113
and Grant-in-Aid for Scientific Research(C) JSPS KAKENHI Grant Number 22K03629.
The work of SHL was supported by IBS under the project code, IBS-R018-D1. 
The work of SHL was also partly supported by the DOE under Award Number DOE-SC0010008.
This work was also performed in part at Aspen Center for Physics, which is supported by National Science Foundation grant PHY-2210452. 
The authors acknowledge the Office of Advanced Research Computing (OARC) at Rutgers, The State University of New Jersey for providing access to the Amarel cluster and associated research computing resources that have contributed to the results reported here (URL: \url{https://oarc.rutgers.edu}).

\appendix
\section{Error Estimate}
\label{subsec: boot}

This appendix quantitatively evaluates the statistical uncertainties in the reweighting results for HW and PY distributions using the bootstrap method.  
The bootstrap method is widely used for estimating errors caused by statistical fluctuations in training data.  
It involves resampling the data with replacement to create multiple independent training sets.  
To estimate the error, ParT model for top vs.~QCD test classification and ParT and AM models for HW reweighting are trained for 10 independent bootstrap models.  Histogram ratios $r( \vert \HW\rightarrow \PY)=N(s\vert \HW; w)/ N(s \vert \PY)$ are calculated for testing samples and all test classifier $\times$ weight combinations.  The average of the ratios and the standard deviations of the $10 \times 10=100$  combinations are computed for each bin. The calculated standard deviation corresponds to the systematical error of the top vs.~QCD classifier and the reweighting due to the statistical fluctuation of the training samples and the initialization of the trainable parameters.  

\begin{figure}[ht!]
\begin{center}
 \begin{minipage}[t]{0.45\linewidth}
\includegraphics[width=\linewidth]{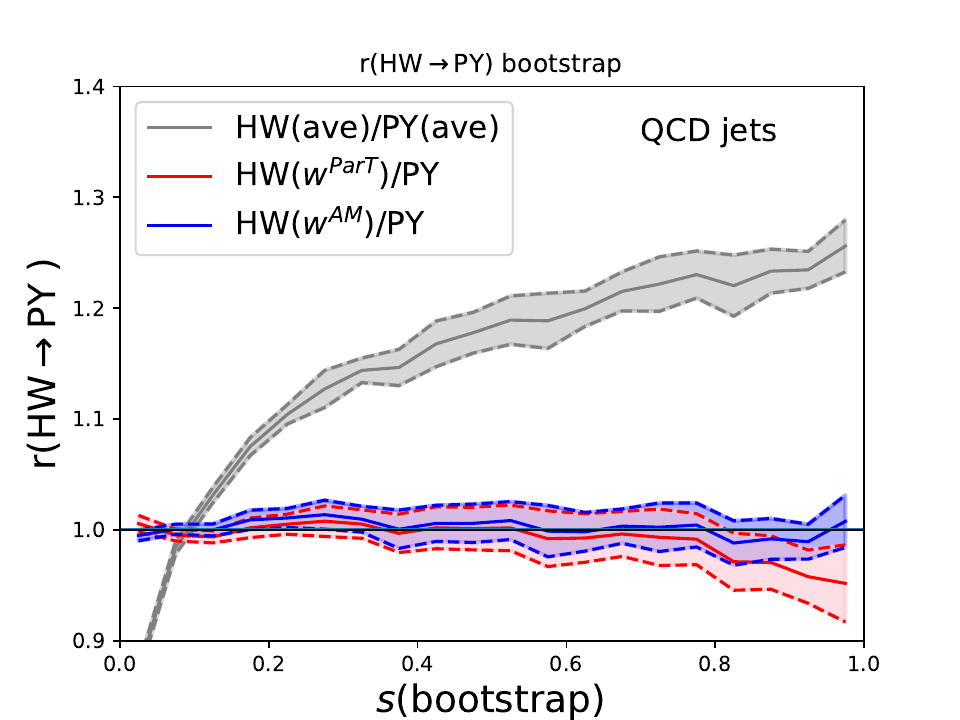}
\subcaption{ $r(\HW\rightarrow\PY)$ for QCD jets }
\end{minipage}
\hskip 0.04\linewidth 
 \begin{minipage}[t]{0.45\linewidth}
\includegraphics[width=\linewidth]{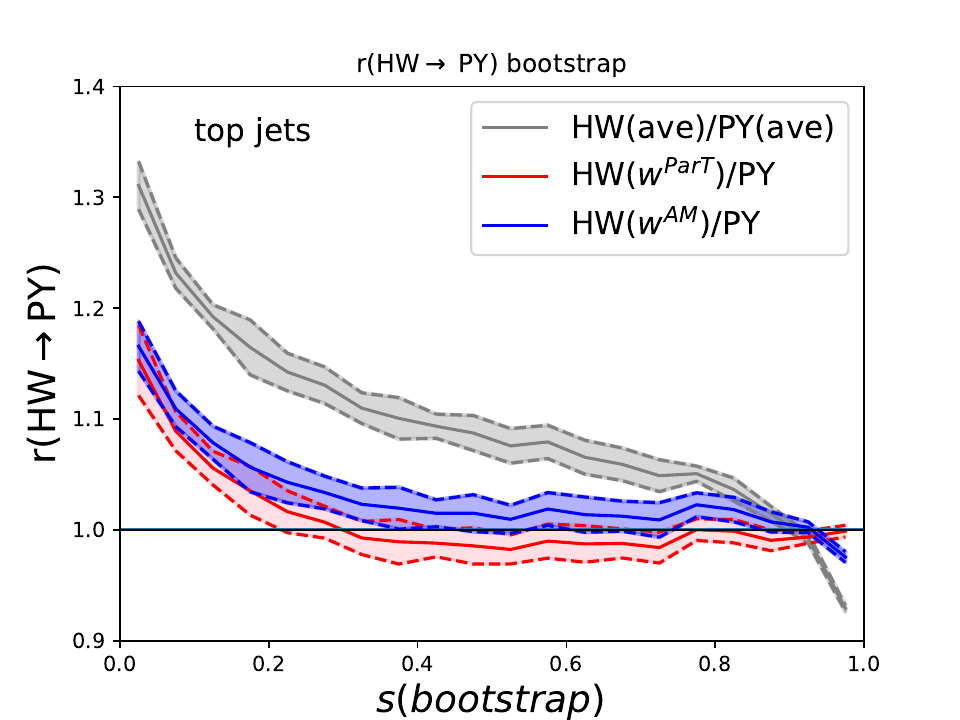}
\subcaption{$r(\HW\rightarrow\PY)$ for top jets }
\end{minipage}
\end{center}
\caption{
The accuracy of the rewriting of HW events distribution to PY event distribution was estimated using the bootstrap method.  
The histogram ratios $r(\HW\rightarrow\PY)$ for top vs.~QCD classification scores are calculated for combinations of top vs.~QCD classifiers and weights trained for 10 independent bootstrap samples for each, and the average and the standard deviations are shown. 
The red (blue) band is the result for $\ParT$ (AM), respectively.
The grey band shows the histogram ratio before reweighting. 
}
\label{fig:sdis}
\end{figure}

The result is given in \figref{fig:sdis}.  
The grey band shows the average values and the std of the HW/PY before the reweighting, while the red (blue) band shows the average and fluctuation for ParT (AM) reweighting, respectively. 
For the QCD jet distribution (a), the blue band (AM) is narrower than the red band (ParT), where the AM reweighting is more aligned to 1.  
Conversely, for the top sample distribution (b), ParT and AM reweightings are away from 1 in QCD-like top jes ($s\sim 0$), highlighting reweighting challenges in these regions. 

Note that we do not bootstrap the testing events; therefore, we underestimate the fluctuation of the testing sample. In addition, we are using the normalized weight $w$ to reduce the training fluctuation, and certain biases in the tail region might be introduced for ParT. 

\bibliographystyle{JHEP}
\bibliography{QCD_MF}

\end{document}